\documentclass[a4paper,fleqn,usenatbib,usedcolumn]{mnras}

\usepackage[T1]{fontenc}
\usepackage{ae,aecompl}

\usepackage{graphicx}	
\usepackage{amsmath}	
\usepackage{amssymb}	

\usepackage{threeparttablex} 
\usepackage{pdflscape} 
\usepackage{longtable}

\defcitealias{MartinezVazquez2015}{Paper~I}
\defcitealias{MartinezVazquez2016}{Paper~II}

\newcommand{\bb}{\hbox{\it B\/}}
\newcommand{\vv}{\hbox{\it V\/}}

\newcommand{\ii}{\hbox{\it I\/}}

\newcommand{\bmv}{\hbox{\bb--\vv\/}}
\newcommand{\bmi}{\hbox{\bb--\ii\/}}

\newcommand{\logp}{\hbox{$\log P$}}
\newcommand{\feh}{\hbox{\feh}}

\title[]{Variable stars in Local Group Galaxies - II. Sculptor dSph}

\author[C. E. Mart{\'i}nez-V{\'a}zquez et al.]{
C. E. Mart{\'i}nez-V{\'a}zquez$^{1,2}$,\thanks{E-mail: clara.marvaz@gmail.com (CEM-V)}
P. B. Stetson$^{3}$, 
M. Monelli$^{1,2}$,
E. J. Bernard$^{4}$, 
G. Fiorentino$^{5}$,
\newauthor{
C. Gallart$^{1,2}$,
G. Bono$^{6,7}$,
S. Cassisi$^{8}$,
M. Dall'Ora$^{9}$,
I. Ferraro$^{7}$,
G. Iannicola$^{7}$, and}
\newauthor{
A. R. Walker$^{10}$}
\\
$^{1}$ Instituto de Astrof{\'i}sica de Canarias (IAC), E-38205 La Laguna, Tenerife, Spain\\ 
$^{2}$ Universidad de La Laguna (ULL), Dpto. Astrof{\'i}sica, E-38206 La Laguna, Tenerife, Spain\\ 
$^{3}$ Herzberg Astronomy and Astrophysics, National Research Council Canada, 5071 West Saanich Road, Victoria, BC V9E 2E7, Canada\\
$^{4}$ Laboratoire Lagrange, Observatoire de la C\^ote d'Azur, 06304 Nice Cedex 4, France\\
$^{5}$ INAF-Osservatorio Astronomico di Bologna, via Ranzani 1, 40127, Bologna, Italy\\
$^{6}$ Dipartimento di Fisica, Universit\`a di Roma Tor Vergata, Via della Ricerca Scientifca 1, 00133 Roma, Italy\\
$^{7}$ INAF-Osservatorio Astronomico di Roma, Via Frascati 33, 00040 Monteporzio Catone, Italy\\
$^{8}$ INAF-Osservatorio Astronomico di Teramo, Via M. Maggini, 64100 Teramo, Italy\\
$^{9}$ INAF-Osservatorio Astronomico di Capodimonte, Via Moiariello 16, 80131 Napoli, Italy\\
$^{10}$ Cerro Tololo Inter-American Observatory, National Optical Astronomy Observatory, Casilla 603, La
Serena, Chile}

\date{Accepted 2016 July 28. Received 2016 July 28; in original form 2016 March 31}

\pubyear{2016}

\begin{document}
\label{firstpage}
\pagerange{\pageref{firstpage}--\pageref{lastpage}}
\maketitle


\begin{abstract}

We present the identification of 634 variable stars in the Milky Way dSph 
satellite Sculptor based on archival ground-based optical observations spanning 
$\sim$24 years and covering $\sim$ 2.5 deg$^2$. We employed the same 
methodologies as the ``Homogeneous Photometry'' series published by Stetson. 
In particular, we have identified and characterized one of the largest (536) RR~Lyrae 
samples so far in a Milky Way dSph satellite. We have also detected four 
Anomalous Cepheids, 23 SX~Phoenicis stars, five eclipsing binaries, three field 
variable stars, three \textit{peculiar} variable stars located above the horizontal 
branch -- near to the locus of BL~Herculis -- that we are unable to classify properly. 
Additionally we identify 37 Long Period Variables plus 23 probable variable stars, 
for which the current data do not allow us to determine the period. We report positions and 
finding charts for all the variable stars, and basic properties (period, amplitude, 
mean magnitude) and light curves for 574 of them. We discuss the properties of the 
RR Lyrae stars in the Bailey diagram, which supports the coexistence of subpopulations 
with different chemical compositions. We estimate the mean mass of Anomalous Cepheids 
($\sim$1.5M$_{\sun}$) and SX~Phoenicis stars ($\sim$1M$_{\sun}$). We discuss in detail 
the nature of the former. The connections between the properties of the different 
families of variable stars are discussed in the context of the star formation history 
of the Sculptor dSph galaxy.

 \end{abstract}

\begin{keywords}
stars: variables: general -- galaxies: evolution -- galaxies: individual: Sculptor dSph -- Local Group -- galaxies: stellar content
\end{keywords}



\section{Introduction}\label{sec:introduction}  

Pulsating variable stars are powerful tools to investigate the evolution of their
host galaxy, as they trace the age and the metallicity of the parent population.
Most importantly, the coexistence of different types of variable stars provides, 
thanks to their pulsational properties, independent constraints not only on the
star formation history and the chemical evolution, but also on the distance of
the system. Indeed, because pulsations occur at specific evolutionary stages that
depend on the stellar mass, variable stars trace the spatial distribution of 
stellar populations of given ages. Therefore they can be used as markers of spatial
trends across the galaxy under examination \citep[e.g.,][]{Gallart2004}. Moreover, 
even the range of pulsational properties among individual stars of a particular 
type can trace some  differences in the age and metallicity of the corresponding 
population \citep[e.g.,][]{Bernard2008,MartinezVazquez2015}. 

This paper focuses on the variable-star content of the Local Group dwarf 
spheroidal (dSph) Sculptor. Sculptor is one of the ``classical'' Milky Way dSph
satellites. After the Magellanic Clouds, it was the first to be discovered along 
with Fornax \citep{Shapley1938}. Sculptor's stellar content has been investigated 
in a large number of papers, using different techniques. Large scale and/or deep
photometric surveys provided colour-magnitude diagrams (CMDs) showing an extended 
horizontal branch \citep[HB,][]{Majewski1999,Hurley-Keller1999,Harbeck2001}, and 
a wide colour spread of the red giant branch first mentioned by \citet{DaCosta1984}. 
While it is well established that Sculptor is composed of a predominantly old population
\citep{Monkiewicz1999,deBoer2011}, it clearly presents some age spread 
\citep{Tolstoy2004,deBoer2012}. The chemical enrichment history of Sculptor has been 
investigated through spectroscopy of its RGB 
\citep{Tolstoy2004,Kirby2009,Starkenburg2013,Skuladottir2015} and HB stars
\citep{Clementini2005}, revealing a large range in metallicity, of the order of
1 dex. In \citet{MartinezVazquez2015} (hereafter \citetalias{MartinezVazquez2015}),
based on the pulsational properties of RR Lyrae (RRL) stars, we showed that a 
similar metallicity spread ($\sim$ 0.8 dex) was already in place at an early 
epoch (>10 Gyr), imprinted in the parent population that we observe today as RRL stars.

The first investigation of the variable-star content in Sculptor dates back to
the work by  \citet{Baade1939} and \citet{ Thackeray1950}. However it was not until
\citet{vanAgt1978} that a conspicuous population of 602 candidate variable
stars was discovered and periods for 64 of them were provided. 

The most complete catalogue of variable stars (in terms of providing
pulsational  properties) in Sculptor is that of \citet{Kaluzny1995}. They
investigated the  central region of the Sculptor dSph ($\sim$ 15$\arcmin
\times$15$\arcmin$) as a side-program  of the OGLE I project. They identified
231 variable stars that were classified as  226 RRL, 3 Anomalous Cepheids (AC), 
and 2 long period variable (LPV) stars. Their
properties are consistent with a metal-poor population ([Fe/H] $<$ --1.7). 
A spectroscopic follow-up made by
\citet{Clementini2005} confirmed this result through low resolution (R$\approx$800) 
spectroscopy of 107 variables using the $\Delta$S
method. In particular, they found that the
metallicity peaks at [Fe/H] $\sim$--1.8.

In \citetalias{MartinezVazquez2015} we reported on the
detection of a large metallicity spread and spatial gradients within the
population of Sculptor's RRL star population. In this work we present the full
catalog of variable stars detected in this galaxy employing the same methodologies 
as the ``Homogeneous Photometry'' series \citep{Stetson1998,Stetson2000,Stetson2003,
Stetson2005a,Stetson2005b,Stetson2014}.
In \S~\ref{sec:photometry}
we present the extensive data set of 4,404 images used in this analysis. In 
\S~\ref{sec:cmd_var} we discuss the variable-star detection and classification. 
We later discuss in detail different families of variable stars: RRL stars 
(\S~\ref{sec:rrl}), AC (\S \ref{sec:acep}), SX Phoenicis (SX~Phe,\S~\ref{sec:sxpho}), 
and other groups (peculiar, binaries, long period and probable variable stars, 
\S~\ref{sec:others}). In \S~\ref{sec:discussion} we discuss the properties of the
old populations of Sculptor, analysing its RRL stars in detail. A summary of our 
conclusions (\S~\ref{sec:conclusions}) closes the paper. We highlight that in the 
online version of the paper we provide full details on all the variable stars 
discussed: time series photometry, light curves, mean photometric and pulsational 
properties and finding charts.


\section{Photometric data set}\label{sec:photometry}

The photometric data set used for this study consists of 5,149 
individual CCD images obtained during 21 observing runs between 1987 
October and 2011 August (i.e., over nearly 24 years). It covers an 
area over the sky of $\sim$4.7 deg$^2$ centred on the Sculptor dSph 
galaxy. However, only 4,404 images (within $\sim$ 2.5 deg$^2$) were
calibrated photometrically, while the area with a significant number 
of epochs for the variability study is further reduced to $\sim$ 
2.0 deg$^2$.
These data were acquired with a variety of cameras on a number of
different telescopes at the European Southern Observatory, Cerro
Tololo Interamerican Observatory, and Las Campanas Observatory as
detailed in the accompanying Table~\ref{tab:obs}.  For each of the observing
runs the table specifies the beginning and ending dates of the run
(although it is not necessarily true that Sculptor was observed on
all the nights during any given run).  Table~\ref{tab:obs} also identifies the
telescope and the detector system used for each of the runs, as
well as the number of separate exposures obtained in the $B$, $V$,
$R$, $I$, and ``other'' filter passbands.  The ``multiplex''
column indicates the number of individual disjoint CCDs in
each instrument.  For instance, the ESO/MPI Wide Field Imager used
during run, ``wfi33'', has eight adjacent CCDs; the six individual
exposures obtained during this run produced $6\times8 = 48$
separate CCD images that contributed to our overall total of
5,149.  However, no individual star was contained in more than six
of the 48 images from that run, since the different CCDs map to
non-overlapping areas on the sky.  Similarly, the SUSI camera on
the ESO NTT telescope had two adjacent CCDs, so the 45 individual
exposures obtained during run 4,``susi9810,'' comprised 90
separate CCD images.  

All 5,149 CCD images were processed to produce instrumental magnitudes
for individual stars using the DAOPHOT/ALLSTAR/ALLFRAME package of
programs \citep[e.g.,][]{Stetson1987,Stetson1994}; those which could be were then calibrated
using the protocols described by \citet{Stetson2000,Stetson2005a}.  These
methods have by now been used in a large number of refereed papers
by members of our collaboration, and more elaborate details are
not needed here.  The few exposures that we have designated as
being obtained with ``other'' filters in Table~\ref{tab:obs} are not
used photometrically here, but they were included in the ALLFRAME
reductions to exploit the information they could provide toward
the completeness of the star list and the quality of the
astrometry.  

\begin{table*}
\begin{scriptsize}
\centering
\caption{Sculptor dSph Observations} 
\label{tab:obs}
  \begin{tabular}{llllcccccccc} 
 \hline 
        Run ID      &   Run Dates             &   Telescope      &  Detector  &   B  &   V  &   R  &   I  &  Other & Multiplex  & Notes \\  
\hline                                                                                                             
		 1	f2          &   1987 Oct 26           &   CTIO 0.9m      &  TI        &  10  &  12  &  --  &  --  &   --   &            &       \\
     2	ogle        &   1993 Jun 20-Sep 05    &   LCO 1.0m       &  FORD2     &  --  &  49  &  --  &  --  &   --   &            &       \\
     3	wfi33       &   1997 Jul 19-23        &   ESO/MPI 2.2m   &  WFI       &  --  &   3  &  --  &   3  &   --   & $\times$8  &       \\
     4	susi9810    &   1998 Oct 26-29        &   ESO NTT 3.6m   &  SUSI      &   8  &  18  &  --  &  19  &   --   & $\times$2  &       \\
     5	susi0008    &   2000 Aug 07-10        &   ESO NTT 3.6m   &  SUSI      &   7  &  --  &  21  &  --  &   --   & $\times$2  &       \\
     6	wfi18       &   2000 Oct 18-22        &   ESO/MPI 2.2m   &  WFI       &   5  &   3  &  --  &   3  &    1   & $\times$8  & a     \\
     7	wfi36       &   2002 Aug 30-Sep 03    &   ESO/MPI 2.2m   &  WFI       &  13  &  19  &  13  &  --  &   --   & $\times$8  & b     \\
     8	wfi34       &   2002 Oct 12-16        &   ESO/MPI 2.2m   &  WFI       &  20  &  14  &  --  &  --  &    2   & $\times$8  & a     \\
     9	wfi35       &   2002 Dec 12-28        &   ESO/MPI 2.2m   &  WFI       &  12  &  12  &  --  &  --  &   --   & $\times$8  &       \\
    10	wfi31       &   2003 Sep 18-20        &   ESO/MPI 2.2m   &  WFI       &   2  &  29  &  --  &  30  &   --   & $\times$8  &       \\
    11	susi0410    &   2004 Oct 05-09        &   ESO NTT 3.6m   &  SUSI      &  --  &  --  &  29  &  --  &   --   & $\times$2  &       \\
    12	susi0709    &   2007 Sep 09-13        &   ESO NTT 3.6m   &  SUSI      &  --  &  --  &  92  &  --  &   --   & $\times$2  &       \\
    13	susi0711    &   2007 Nov 07-09        &   ESO NTT 3.6m   &  SUSI      &  --  &  --  &   6  &  --  &   --   & $\times$2  &       \\
    14	ct08nov     &   2008 Jul 18-19        &   CTIO 0.9m      &  Tek2K\_3  &  15  &  15  &  --  &  15  &        &            &       \\
    15	B0809       &   2008 Sep 06-07        &   CTIO 4.0m      &  Mosaic2   &  20  &  27  &  --  &  12  &   --   & $\times$8  &       \\
    16	B08sep      &   2008 Sep 24-27        &   CTIO 4.0m      &  Mosaic2   &  98  &  98  &  --  &  14  &   --   & $\times$8  &       \\
    17	susi0809    &   2008 Sep 28-29        &   ESO NTT 3.6m   &  SUSI      &  10  &  --  &  20  &  --  &   --   & $\times$2  &       \\
    18	B0911       &   2009 Nov 20-24        &   CTIO 4.0m      &  Mosaic2   &  11  &  39  &  --  &  47  &   --   & $\times$8  &       \\
    19	soar1010    &   2010 Oct 14           &   SOAR 4.1m      &  SOI       &  --  &  59  &  --  &  --  &   --   & $\times$2  &       \\
    20	soar1012    &   2010 Dec 04           &   SOAR 4.1m      &  SOI       &  --  &  10  &  --  &  --  &   --   & $\times$2  &       \\
    21	lee1        &   2011 Aug 30           &   CTIO 4.0m      &  Mosaic2   &   3  &   3  &  --  &  --  &    3   & $\times$7  & c     \\
		\hline
		\end{tabular}
		\begin{tablenotes}
		\item Notes. --
(a) ``Other'' filter was $U$; these images were not photometrically calibrated.
(b) ``V'' exposures comprised 9 exposures in the DDO 51 filter, plus 10 exposures in the Washington M filter.
These were approximately transformed to the Landolt $V$ system.
(c)  Observations were made in the Str\"omgren $b$ and $y$ filters, and a Calcium H+K filter.  The instrumental
$b$ and $y$ magnitudes were approximately transformed to Landolt $B$ and $V$; the Ca measurements were not
calibrated.

\item Data credits. --
 (1)  Data contributed by N.~B.~Suntzeff;
 (2)  Data contributed by A.~Udalski --OGLE project--;
 (3)  Program ID unknown, observer unknown;
 (4)  Program ID 62.N-0653(A), observer unknown;
 (5)  Program ID 65.N-0472(A), observer unknown;
 (6)  Program ID 066.B-0615, observer L.~Rizzi;
 (7)  Program ID 60.A-9121(A), observer unknown;
 (8)  Program ID 70.B-0696(A), observer unknown;
 (9)  Program ID 70.B-0696(A), observer unknown;
(10)  Program ID 171.B-0588(C), observer unknown;
(11)  Program ID 074.B-0456(A), obsever unknown;
(12)  Program ID 079.B-0379(A), observer unknown;
(13)  Program ID 080.B-0144(A), observer unknown;
(14)  Observer J.~Vasquez, comment ``Saha'';
(15)  Proposal ID 2008B-0143, proposer Saha, observers Saha, Tolstoy, de~Boer;
(16)  Proposal ID 2008B-0206, proposer Bernard, observers Bernard, Walker;
(17)  Program ID 081.B-0534(A), observer unknown;
(18)  Proposal ID 2009B-0157, proposer Saha, observers Olsen, de~Boer;
(19)  Proposal ``Sculptor Optical Imaging,'' proposer Zepf, observers Zepf, Peacock;
(20)  Proposal ID 2010B-0415, proposer H.~Smith, observer unknown;
(21)  Proposal ID 2011B-0139, proposer J.-W.~Lee, observer J.-W.~Lee.
\end{tablenotes}
\end{scriptsize}
\end{table*}

From two of the observing runs, 7 = wfi36 and 21 = lee1, we
inferred at least some magnitudes on the standard Landolt
photometric system from observations that were obtained in
non-standard filters.
In particular, in run 7 (wfi36) 13 exposures were obtained using
the standard Wide-Field Imager ``B'' filter and 13 were obtained
with the standard ``R'' filter.  However, an additional nine
exposures were obtained using a DDO51 (magnesium hydride) filter
(MB\#516/16\_ESO871, central wavelength 516.5$\,$nm, width
16.2$\,$nm) and another ten were obtained with a Washington $M$
filter.  (In a quick search on the internet we were unable to
locate the filter properties for the ESO/WFI $M$ filter, but the
defining $M$ filter has central wavelength 508.5$\,$nm and width
105.0$\,$nm: \citealt{Canterna1976}) By simple trial we
learned that, allowing for quadratic colour terms in the
calibrating transformation equations, we were able to use these
filters to predict Landolt-system $V$-band magnitudes with a
star-to-star reliability $\sigma \sim 0.05\,$mag from the DDO51
magnitudes, and $\sim 0.02\,$mag from the $M$ magnitudes.  In our
opinion, this precision is good enough to use these observations
to help define the variable-star light curves, but we would
hesitate to use these in calibrating our best CMD for Sculptor.

Similarly, in the run we have identified as ``lee1'' (our number
21), the observations were made in the Str\"omgren $b$ and $y$
filters, and in a Calcium H+K filter.  (We were unable to locate
the filter characteristics of the Ca filter then in use with the CTIO
4.0m Mosaic Camera, but the filter used on the smaller CTIO
telescopes is described as having central wavelength 396.0$\,$nm and
width 10.0$\,$nm.)  We made no attempt to calibrate the Ca
observations photometrically, but we found that with quadratic
colour coefficients $b$ could be transformed to $B$ with a
star-to-star reliability of 0.03$\,$mag, and $y$ transformed to
$V$ with a reliability of 0.02$\,$mag.  Again, we feel that these
are good enough to employ for the variable-star light curves, but
would hesitate to include them in the CMD.

The individual $B$, $V$ and $I$ measurements for all of the detected 
variables in the calibrated field of Sculptor are listed in 
Table~\ref{tab:photometry}. They were named with the prefix ``scl-CEMV'' 
(which refers to the name of the galaxy and the current work), followed 
by a number which increases in order of increasing right ascension.

\begin{table*}
\begin{scriptsize}
\centering
\caption{Photometry of the variable stars in Sculptor dSph.} 
\label{tab:photometry}
  \begin{tabular}{ccccccccc} 
 \hline
MHJD$^a$    &    $B$    &    $\sigma_B$    &    MHJD$^a$    &    $V$    &    $\sigma_V$    &    MHJD$^a$    &    $I$    &    $\sigma_I$   \\
\hline
\multicolumn{9}{c}{scl-CEMV001} \\
\hline
   54734.613281  &   19.636  &    0.011  &   54734.609375  &   19.429  &    0.014  &   54734.875000  &   19.460  &    0.014 \\
   54734.656250  &   19.809  &    0.017  &   54734.660156  &   19.592  &    0.013  &   54734.898438  &   19.475  &    0.038 \\
   54734.718750  &   20.070  &    0.026  &   54734.710938  &   19.800  &    0.026  &   54734.902344  &   19.403  &    0.034 \\
   54734.750000  &   20.314  &    0.311  &   54734.777344  &   19.920  &    0.010  &   54734.914062  &   19.806  &    0.363 \\
   54734.773438  &   20.220  &    0.010  &   54734.828125  &   20.006  &    0.009  &   52902.726562  &   19.512  &    0.039 \\
   54734.824219  &   20.369  &    0.011  &   54735.699219  &   20.238  &    0.016  &   52902.730469  &   19.450  &    0.058 \\
   54735.707031  &   20.681  &    0.014  &   54736.550781  &   20.371  &    0.020  &   ---           &   ---     &    ---   \\
   54736.554688  &   20.794  &    0.020  &   54736.742188  &   19.536  &    0.010  &   ---           &   ---     &    ---   \\
   54736.746094  &   19.746  &    0.010  &   54736.816406  &   19.732  &    0.012  &   ---           &   ---     &    ---   \\
   54736.824219  &   20.058  &    0.012  &   54736.859375  &   19.893  &    0.018  &   ---           &   ---     &    ---   \\
   54736.859375  &   20.150  &    0.015  &   54736.894531  &   19.921  &    0.020  &   ---           &   ---     &    ---   \\
   54736.898438  &   20.480  &    0.038  &   52902.714844  &   20.041  &    0.036  &   ---           &   ---     &    ---   \\
   ---           &   ---     &    ---    &   52902.722656  &   20.050  &    0.025  &   ---           &   ---     &    ---   \\
\hline
\multicolumn{9}{c}{scl-CEMV002} \\
\hline
   54716.597656  &   20.207  &    0.028  &   54716.613281  &   20.060  &    0.029  &   54716.632812  &   19.757  &    0.068 \\
   54716.601562  &   20.228  &    0.013  &   54716.617188  &   20.002  &    0.017  &   54716.632812  &   19.841  &    0.025 \\
   54716.609375  &   20.194  &    0.011  &   54716.621094  &   20.068  &    0.008  &   54716.640625  &   19.807  &    0.010 \\
   54716.613281  &   20.199  &    0.013  &   54716.625000  &   20.079  &    0.009  &   54716.648438  &   19.761  &    0.007 \\
   54734.566406  &   20.384  &    0.010  &   54716.628906  &   20.072  &    0.009  &   54716.656250  &   19.791  &    0.007 \\
   54734.613281  &   20.210  &    0.009  &   54734.558594  &   20.233  &    0.011  &   54734.875000  &   19.842  &    0.014 \\
   54734.656250  &   20.234  &    0.009  &   54734.609375  &   20.071  &    0.011  &   54734.902344  &   19.832  &    0.045 \\
   54734.773438  &   20.425  &    0.010  &   54734.660156  &   20.082  &    0.008  &   55157.625000  &   19.880  &    0.009 \\
   54734.824219  &   20.487  &    0.012  &   54734.777344  &   20.241  &    0.008  &   55157.632812  &   19.888  &    0.008 \\
   54735.554688  &   20.200  &    0.012  &   54734.828125  &   20.294  &    0.009  &   55157.640625  &   19.924  &    0.008 \\
   54735.605469  &   20.218  &    0.011  &   54735.546875  &   20.035  &    0.013  &   55157.648438  &   19.942  &    0.009 \\
   54735.652344  &   20.273  &    0.011  &   54735.597656  &   20.062  &    0.012  &   52902.726562  &   19.904  &    0.038 \\
   54735.707031  &   20.457  &    0.011  &   54735.656250  &   20.137  &    0.014  &   ---           &   ---     &    ---   \\
   54735.753906  &   20.483  &    0.013  &   54735.699219  &   20.289  &    0.013  &   ---           &   ---     &    ---   \\
   54736.554688  &   20.238  &    0.018  &   54735.746094  &   20.277  &    0.014  &   ---           &   ---     &    ---   \\
   54736.601562  &   20.383  &    0.014  &   54736.550781  &   20.121  &    0.019  &   ---           &   ---     &    ---   \\
   54736.640625  &   20.517  &    0.013  &   54736.601562  &   20.238  &    0.015  &   ---           &   ---     &    ---   \\
   54736.707031  &   20.424  &    0.015  &   54736.636719  &   20.293  &    0.018  &   ---           &   ---     &    ---   \\
   54736.746094  &   20.190  &    0.012  &   54736.710938  &   20.190  &    0.021  &   ---           &   ---     &    ---   \\
   54736.781250  &   20.205  &    0.013  &   54736.742188  &   20.100  &    0.016  &   ---           &   ---     &    ---   \\
   54736.824219  &   20.203  &    0.011  &   54736.785156  &   20.029  &    0.014  &   ---           &   ---     &    ---   \\
   54736.859375  &   20.247  &    0.013  &   54736.816406  &   20.014  &    0.015  &   ---           &   ---     &    ---   \\
   54736.898438  &   20.334  &    0.040  &   54736.859375  &   20.098  &    0.017  &   ---           &   ---     &    ---   \\
   55159.722656  &   20.339  &    0.010  &   54736.894531  &   20.239  &    0.023  &   ---           &   ---     &    ---   \\
   55159.726562  &   20.350  &    0.009  &   55157.597656  &   20.172  &    0.007  &   ---           &   ---     &    ---   \\
   52517.843750  &   20.263  &    0.010  &   55157.613281  &   20.214  &    0.007  &   ---           &   ---     &    ---   \\
   52517.851562  &   20.171  &    0.009  &   52902.714844  &   20.318  &    0.013  &   ---           &   ---     &    ---   \\
   52517.855469  &   20.212  &    0.010  &   52518.730469  &   20.242  &    0.025  &   ---           &   ---     &    ---   \\
   ---           &   ---     &    ---    &   52518.730469  &   20.213  &    0.023  &   ---           &   ---     &    ---   \\
   ---           &   ---     &    ---    &   52518.734375  &   20.294  &    0.023  &   ---           &   ---     &    ---   \\
   ---           &   ---     &    ---    &   52518.742188  &   20.104  &    0.038  &   ---           &   ---     &    ---   \\
   ---           &   ---     &    ---    &   52518.746094  &   20.185  &    0.037  &   ---           &   ---     &    ---   \\
   ---           &   ---     &    ---    &   52518.753906  &   20.123  &    0.039  &   ---           &   ---     &    ---   \\
\hline
\end{tabular}
\begin{tablenotes}
\item $^a$ Modified Heliocentric Julian Date of mid-exposure: HJD - 2,400,000
\item (This table is a portion of its entirely form which will be available in the online journal.)
\end{tablenotes}
\end{scriptsize}
\end{table*}

\section{Variable star identification and Colour-magnitude diagram}\label{sec:cmd_var}

\begin{figure}
\includegraphics[scale=0.5]{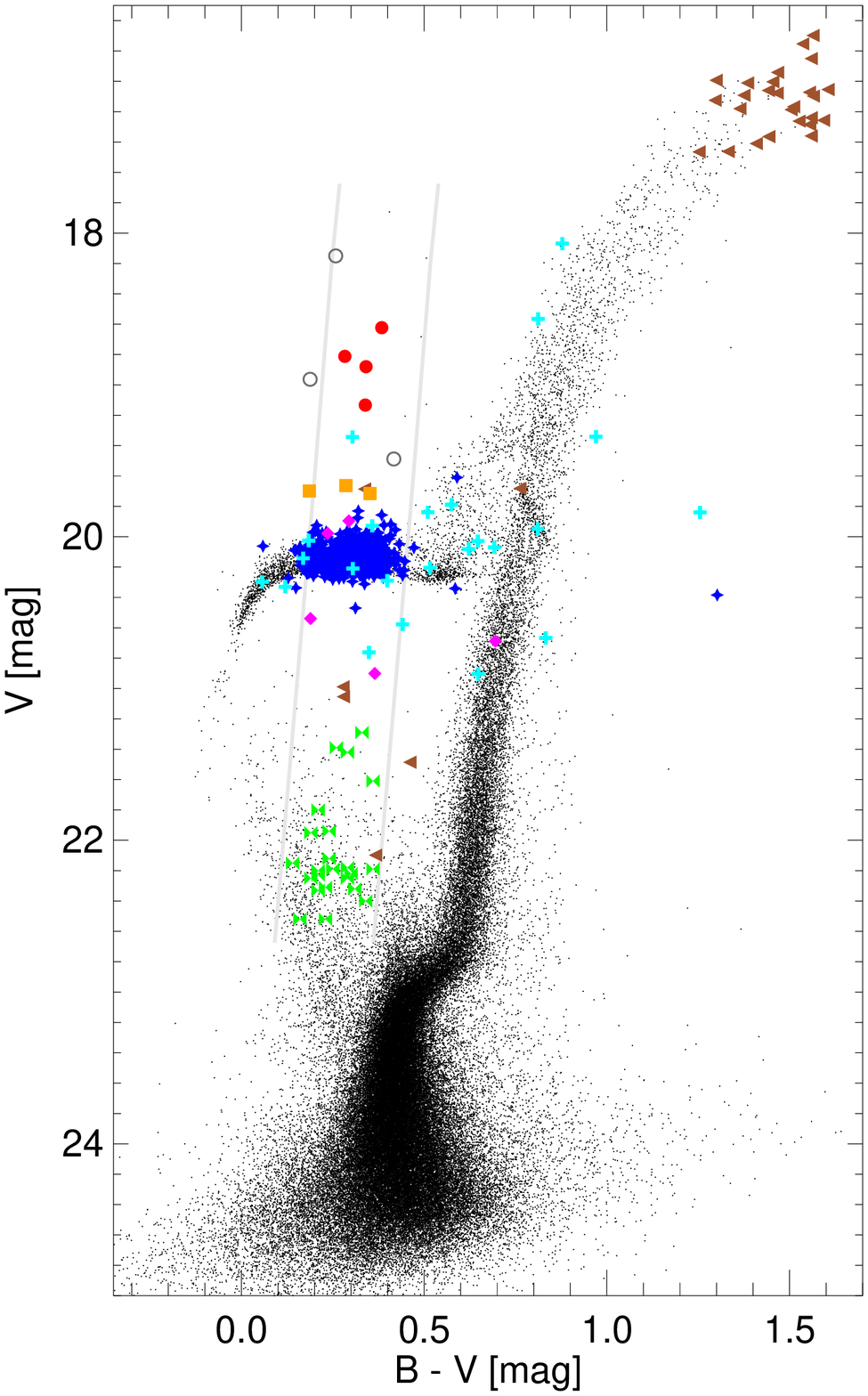}
\caption{Colour-magnitude diagram of Sculptor with the identification of the all 
detected variable stars. Green bowtie symbols represent the SX Phoenicis stars. 
Magenta diamonds are the probable eclipsing binaries.  Blue stars are the RR 
Lyrae stars. Orange square are the three peculiar variables detected in 
Sculptor. Red circles are anomalous cepheids. Grey open circles are the field 
variables detected in Sculptor. Brown triangles are the long period variable 
stars. Large cyan plusses are probable variables. 
The edges of the Instability Strip are those presented in \citet{Fiorentino2006} 
extended to low luminosities (light grey lines).}
\label{fig:cmd}
\end{figure}

We performed the variability search over the full data set, using an updated
version of the Welch-Stetson variability index \citep{Welch1993,Stetson1996}
which identifies candidate variable stars on the basis of our multi-band
photometry. From the list of 663 variable candidates, we have identified 
574 as actual variable stars (i.e., we can derive periods, amplitudes 
and mean magnitudes in $BVI$), 60 as likely variable stars, and 29 as 
non-variable stars. Figure \ref{fig:cmd} presents the (\vv~, \bmv) CMD of 
Sculptor, with the detected variable stars highlighted. 
Most of them belong to Sculptor and are located in the instability strip (IS), 
spanning a wide range of luminosities.
In particular, from the brightest to the faintest, we identified four ACs (red
circles), three \textit{peculiar} HB variable stars (orange squares, similar
variables were detected in \citealt{Coppola2013}), 536 RRL stars (blue star 
symbols), five eclipsing binaries (magenta diamonds) and 23 SX~Phe stars 
(green bowties). Moreover, a sample of 31 probable LPV stars (brown triangles)
are found near the tip of the Sculptor's giant branch, plus six more LPVs are 
spread over the CMD. The three grey open circles mark the position of variable 
stars that we believe are located along the line of sight of the Sculptor dSph, 
two of them compatible with being field $\delta$~Scuti while the other one is 
a possible field RRc. Finally, large cyan plusses are 23 stars identified
as probable variables. Table~\ref{tab:summary} reports a summary of the full 
list of putative variable stars.


\begin{table*}
\begin{scriptsize}
\centering
\caption{Summary of the detected variable stars inside $\sim$ 2 deg$^2$ centred on Sculptor dSph.} 
\label{tab:summary}
  \begin{tabular}{lccccc}
 \hline
Type of variable & Total & Fundamental Mode & First Overtone & Second Overtone & Double Mode \\
\hline 
\textbf{ACTUAL} & & & & & \\
\hline
AC                           &   4    &   4   &   0    &   0    &  ---  \\ 
RRL                          &  536   &  289  &  197   &  ---   &  50   \\ 
SX~Phe                       &  23    &   ?$^{*}$   &  ?$^{*}$  &  ?$^{*}$  &  ---   \\ 
Eclipsing binary             &  5     &  ---  &  ---   &  ---   &  ---  \\ 
Field                        &  2$^{a}$ + 1$^{b}$    &  ---  &  ---   &  ---   &  ---  \\ 
\textit{Peculiar} HB         &  3$^{c}$    &  ---  &  ---   &  ---   &  ---  \\ 
\hline
\textbf{LIKELY} & & & & & \\
\hline
LPV                          & 31$^{d}$ + 6$^{e}$   &  ---  &  ---   &  ---   &  ---  \\ 
Probable$^{f}$               & 23   &  ---  &  ---   &  ---   &  ---  \\ 
\hline
\end{tabular}
\begin{tablenotes}
\item $^{*}$ See \S~\ref{sec:sxpho} for a detailed discussion of the classification mode.
\item $^{a}$ Compatible with being field $\delta$~Scuti.
\item $^{b}$ Compatible with being field RRc.
\item $^{c}$ In \S~\ref{sec:peculiar}, the reader have the explanation of why these stars are
considered \textit{peculiars}.  
\item $^{d}$ LPV stars near the tip of the red-giant branch.
\item $^{e}$ LPV stars out of the tip of the red-giant branch.
\item $^{f}$ Variable stars for which a proper light curve and a reliable classification is difficult to obtain. 
\end{tablenotes}
\end{scriptsize}
\end{table*}

We derived pulsational properties for all the variable stars using our
\bb\vv\ii-Johnson/Cousins photometry. An initial period search was carried out
using a simple string-length algorithm \citep{Stetson1998b}. The
intensity-averaged magnitudes and amplitudes of the mono-periodic light curves
were obtained by fitting the light curves with a set of templates partly based
on the set of \citet{Layden1999} following the method described in
\citet{Bernard2009}.

\begin{figure*}
	\includegraphics[scale=0.9]{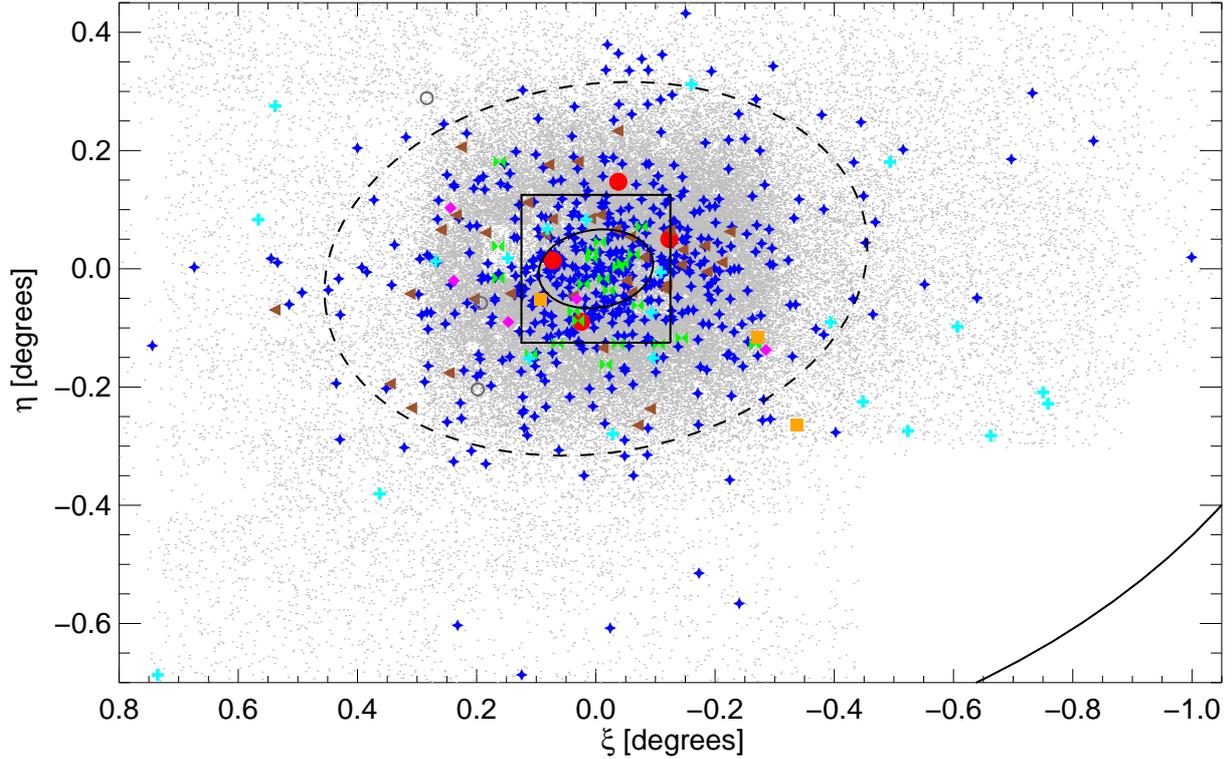}
\caption{Spatial distribution of the detected variable stars in and around 
the Sculptor dSph based our current photometry database. Static stars are 
represented by gray dots. The RRL stars, AC, SX~Phe, eclipsing binary, field 
variable stars, LPV, the three peculiar variables and the probable variable
are shown, with the same symbols as in Fig.~\ref{fig:cmd}. 
The innermost ellipse represents the core ($r_c$=5.8 arcmin; \citealt{Mateo1998}). 
The outermost ellipse (of which only a small arc appear in the south-west corner) 
corresponds to the tidal radius ($r_t$=76.5 arcmin; \citealt{Irwin1995}). 
The dashed ellipse is the radius ($\sim$ 27.5 arcmin) from which the number of 
points per light curve of RRL is lower than 40 in the \bb~ and \vv~ bands. 
The field of view of the study presented in \citet{Kaluzny1995} 
(15\arcmin $\times$ 15\arcmin) is represented by the inner square. 
This field ($\sim$ 2 deg$^2$) covers the area where \citet{vanAgt1978}
detected the 95 percents of his candidates.}
\label{fig:spatial}
\end{figure*}

Fig.~\ref{fig:spatial} presents the spatial distribution of the detected
variable stars (inside of $\sim$2 deg$^2$), superimposed on the 
photometrically calibrated part of the field. The solid ellipses mark the core and tidal radii 
(the latter is only partially visible in the south-west corner).  
As the present database consists of a large selection
of observations collected from different projects, the number of phase points
is not constant over the field and increases towards the centre of
Sculptor. The dashed ellipse marks the area corresponding to the elliptical
radius (equivalent  distance along the semi-major axis) of 27.5$\arcmin$ 
where we estimate the completeness to be homogeneous (discussed below)
in the variable star detection, at least in the magnitude and period range typical of RRL stars. 
The inner square marks the area covered by the \citet{Kaluzny1995} RRL stars. 
It is worth mentioning that the covered area is nearly similar to that observed 
by \citet{vanAgt1978}. He identified 95 percent of his candidates (602) inside
of our current area. However, \citet{vanAgt1978} provided periods only for a 
few (64). On the other hand, we note that, given the large tidal radius of 
Sculptor, many RRL stars are likely still to be discovered. 


\begin{figure}
\hspace{-1cm}
\includegraphics[scale=0.6]{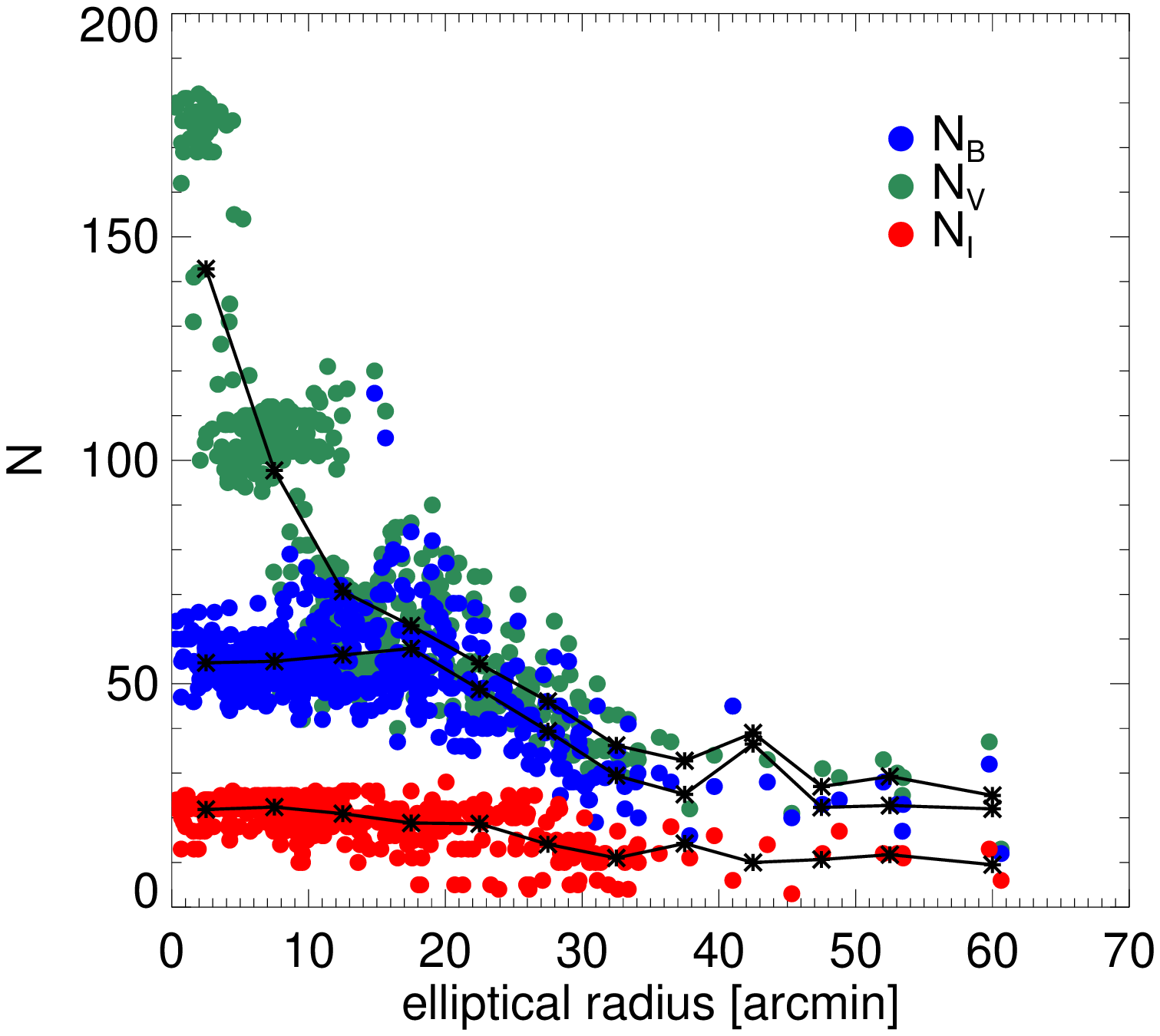}
\caption{Number of phase points for each identified RRL star as a function 
of the elliptical radius for the \bb~ (blue dots), \vv~ (green dots), and \ii~ 
(red dots) filters. Asterisks represent the mean value of each of them every 
5\arcmin (with exception of the last one that covers 10\arcmin, from 55\arcmin 
to 65\arcmin).}
\label{fig:number}
\end{figure}

Focusing on the homogeneity of the sample, Fig.~\ref{fig:number} 
shows the number of phase points for each identified RRL
star as a function of the elliptical radius for the \bb, \vv, and \ii~filters
(blue, green, and red dots, respectively). The plot shows that the \vv~band has
the largest number of points in the central regions (greater than 170 for
elliptical radius $<5\arcmin$ and lower than 100 for elliptical radius $>5\arcmin$ and
$<20\arcmin$), while the \bb~ observations
number above 50 out to $\sim$20$\arcmin$. The mean number of points
in both bands remains above 40 out to 27.5$\arcmin$. The number of \ii~band
points is relatively constant ($\sim$ 15) out to the same  distance and then
slowly declines. Given the large number of independent observing runs and the
large time baseline, we are confident that we have a high and relatively homogeneous
completeness for detection of RRL stars out to 27.5$\arcmin$. This is shown as a
dashed ellipse in Fig.~\ref{fig:spatial}.

\section{RR Lyrae stars}\label{sec:rrl}

\subsection{Classification}\label{sec:classification_rrl}

Based on the pulsational properties, light curves (LCs), and positions on the   
CMD, we identify 536 RRL stars. Of these, 390 were flagged by   
\citet{vanAgt1978} as candidate variables (but periods were provided for only 
53 of them), and 65 were discovered by \citet{Kaluzny1995}; the remaining 81 
are new discoveries. \citet{Kaluzny1995} presented the analysis of 226 RRLs, 
although we show in \S~\ref{sec:kaluzny_comparation} that 10 of these are 
probably not RRLs. Here we provide the pulsation parameters (period, mean 
magnitude and amplitude) for 320 RRLs for the first time.

We sub-classify the sample of RRL stars as: {\em i)} 289 RRab, pulsating
in the fundamental mode; 20 of them are suspected Blazhko stars,
\citep{Blazhko1907}; {\em ii)} 197 RRc, pulsating in the first-overtone mode;
and {\em iii)} 50 possible multi-mode RRd stars, pulsating in both modes
simultaneously. The classification of the latter was uncertain in some cases due
to their relatively noisy or (very) poor light curves. 

\begin{figure}
\includegraphics[scale=0.5]{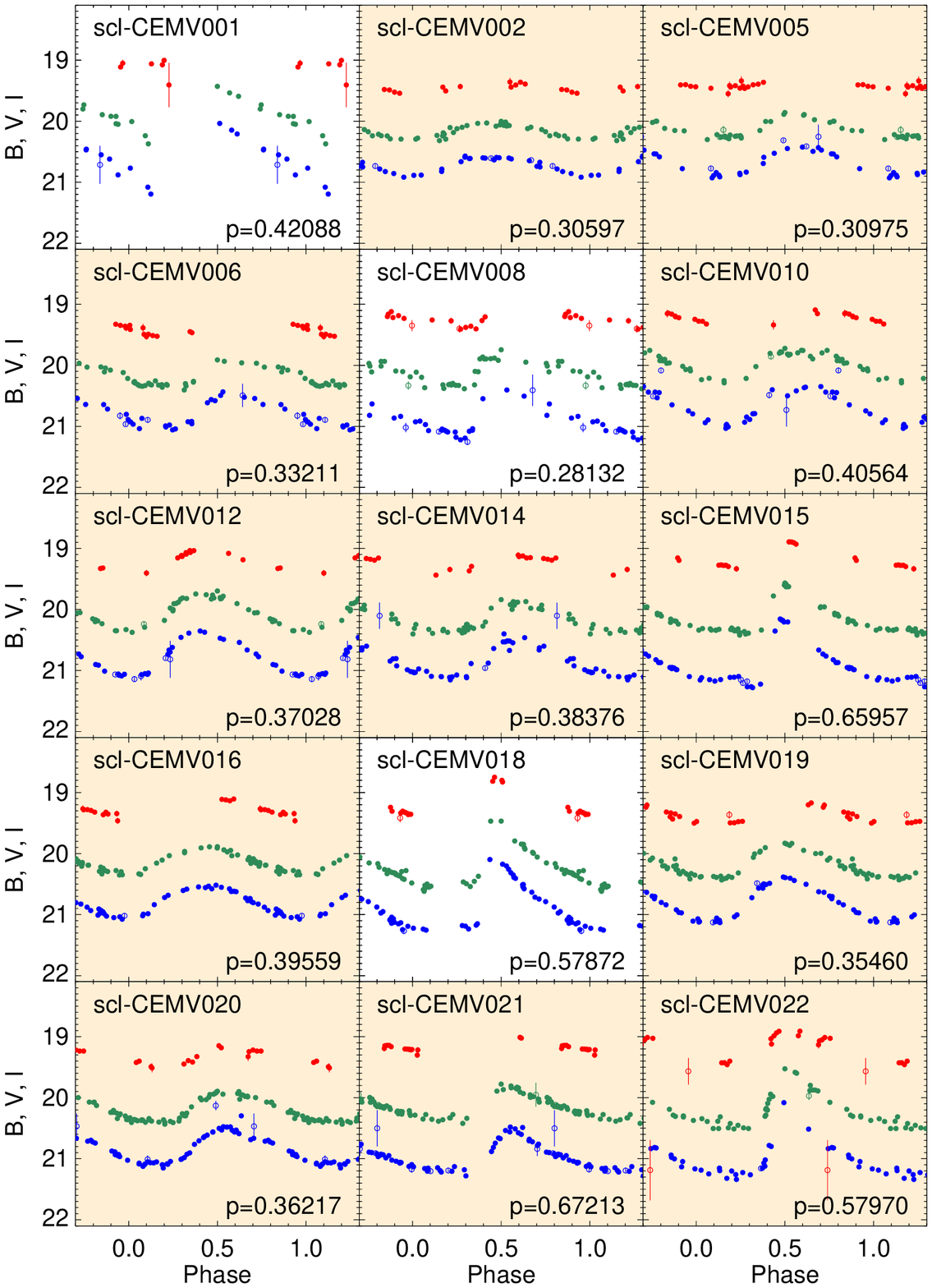}
\caption{Sample of light curves of the RRL stars in the \bb~ (blue), \vv~ 
(green) and \ii~ (red) bands, phased with the period in days given in the 
lower right-hand corner of each panel. The name of the variable is given in 
the left-hand corner of each panel. Open symbols show the bad data points, 
i.e, with errors larger than 3$\sigma$ above the mean error of a given star; 
these were not used in the calculation of the period and mean magnitudes. 
For clarity, the \bb~ and \ii~ light curves have been shifted by 0.4 mag 
down- and upward, respectively. 
Tinted backgrounds mark those RRL stars classified as members of the 
\textit{clean} sample. All light curves are available as Supporting Information 
with the online version of the paper.}
\label{fig:rrl_lcv}
\end{figure}

The LCs of all the RRL stars are presented in Fig.~\ref{fig:rrl_lcv} and their
basic properties (position, period, amplitude and mean magnitude in \bb, \vv~
and \ii~ Johnson/Cousins bands) are detailed in Table~\ref{tab:rrl}. The mean
(maximum) number of points in the light-curves of the RRL stars are 52, 83, and 21
(115, 182 and 28) respectively in \bb, \vv, and \ii. 

\begin{table*}
\begin{scriptsize}
\centering
\caption{Parameters of the RRL stars in Sculptor dSph.} 
\label{tab:rrl}
  \begin{tabular}{lcccccccccccccc } 
 \hline
CEMV+2016 & Original & Alternative & RA & DEC & Period & <\bb> & <\vv> & <\ii> & A$_{\bb}$ & A$_{\vv}$ & A$_{\ii}$ & Q1$^{a}$ & Q2$^{b}$ & Type \\
name & name & name & (J2000) & (J2000) & (current) & & & & & & & & & \\ 
\hline 
scl-CEMV001   &      V461   &   ---   &  00 55 21.02  &  -33 41 01.5  &    0.420876::  &  20.150  & 19.830  & 19.460  &  0.900  &  0.630  &  0.000  &  1  &  X  &  RRab  \\
scl-CEMV002   &      ---	  &   ---   &  00 56 09.15  &  -33 29 20.1  &    0.3059705   &  20.326  & 20.163  & 19.840  &  0.298  &  0.254  &  0.168  &  0  &  1  &  RRc   \\
scl-CEMV005   &      ---	  &   ---   &  00 56 38.94  &  -33 24 32.4  &    0.309746    &  20.260  & 20.086  & 19.778  &  0.494  &  0.359  &  0.165  &  0  &  1  &  RRc   \\
scl-CEMV006   &      ---	  &   ---   &  00 56 48.78  &  -33 31 16.3  &    0.3321113   &  20.333  & 20.117  & 19.738  &  0.568  &  0.479  &  0.401  &  0  &  1  &  RRc   \\
scl-CEMV008   &      ---	  &   ---   &  00 57 04.83  &  -33 45 20.6  &    0.2813182   &  20.357  & 20.047  & 19.570  &  0.881  &  0.576  &  0.485  &  2  &  0  &  RRc   \\
scl-CEMV010   &      ---	  &   ---   &  00 57 30.30  &  -33 44 01.7  &    0.4056391   &  20.252  & 19.982  & 19.637  &  0.632  &  0.466  &  0.265  &  0  &  1  &  RRc   \\
scl-CEMV012   &      V483	  &   ---   &  00 57 41.04  &  -33 30 21.9  &    0.3702771   &  20.307  & 20.037  & 19.619  &  0.740  &  0.613  &  0.424  &  0  &  1  &  RRc   \\
scl-CEMV014   &      V348	  &   ---   &  00 57 54.19  &  -33 37 43.5  &    0.3837578   &  20.405  & 20.131  & 19.676  &  0.606  &  0.469  &  0.276  &  0  &  1  &  RRc   \\
scl-CEMV015   &      V301	  &   ---   &  00 57 55.25  &  -33 47 05.1  &    0.659572    &  20.418  & 20.077  & 19.539  &  1.075  &  0.886  &  0.451  &  0  &  1  &  RRab  \\
scl-CEMV016   &      V302	  &   ---   &  00 57 59.11  &  -33 39 50.6  &    0.3955858   &  20.362  & 20.101  & 19.609  &  0.515  &  0.442  &  0.336  &  0  &  1  &  RRc   \\
scl-CEMV018   &      V363	  &   ---   &  00 57 59.90  &  -33 35 06.7  &    0.5787198   &  20.400  & 20.110  & 19.586  &  1.366  &  1.187  &  0.700  &  0  &  0  &  RRab  \\
scl-CEMV019   &      V535	  &   ---   &  00 58 01.34  &  -33 27 36.1  &    0.3545968   &  20.347  & 20.117  & 19.749  &  0.734  &  0.582  &  0.337  &  0  &  1  &  RRc   \\
scl-CEMV020   &      V482	  &   ---   &  00 58 04.91  &  -33 31 39.9  &    0.3621706   &  20.397  & 20.151  & 19.724  &  0.591  &  0.499  &  0.287  &  0  &  1  &  RRc   \\
scl-CEMV021   &      V520	  &   ---   &  00 58 04.83  &  -33 45 33.3  &    0.6721255   &  20.537  & 20.152  & 19.546  &  0.726  &  0.645  &  0.336  &  0  &  1  &  RRab  \\
scl-CEMV022   &      V326	  &   ---   &  00 58 12.91  &  -33 59 05.8  &    0.5797025   &  20.414  & 20.109  & 19.579  &  1.210  &  0.985  &  0.572  &  0  &  1  &  RRab  \\
scl-CEMV024   &      V334	  &   ---   &  00 58 18.89  &  -33 49 10.9  &    0.537455    &  20.578  & 20.254  & 19.831  &  0.793  &  0.644  &  0.399  &  3  &  0  &  RRab  \\
scl-CEMV025   &      V456	  &   ---   &  00 58 19.14  &  -33 36 29.6  &    0.5928154   &  20.468  & 20.136  & 19.616  &  1.202  &  0.910  &  0.585  &  0  &  1  &  RRab  \\
scl-CEMV026   &      ---	  &   ---   &  00 58 20.51  &  -33 26 54.2  &    0.3934555   &  20.370  & 20.120  & 19.649  &  0.535  &  0.418  &  0.325  &  0  &  1  &  RRc   \\
scl-CEMV027   &      ---	  &   ---   &  00 58 22.85  &  -33 48 34.8  &    0.3700876   &  20.465  & 20.194  & 19.742  &  0.657  &  0.499  &  0.368  &  2  &  0  &  RRc   \\
scl-CEMV028   &      ---	  &   ---   &  00 58 31.22  &  -33 35 27.9  &    0.3203025   &  20.388  & 20.184  & 19.824  &  0.705  &  0.567  &  0.238  &  0  &  1  &  RRc   \\
scl-CEMV030   &      ---	  &   ---   &  00 58 32.76  &  -33 46 08.8  &    0.6897029   &  20.408  & 20.036  & 19.468  &  0.971  &  0.725  &  0.388  &  0  &  1  &  RRab  \\
scl-CEMV031   &      V399	  &   ---   &  00 58 33.54  &  -33 37 23.7  &    0.5836000   &  20.519  & 20.173  & 19.606  &  0.652  &  0.548  &  0.306  &  0  &  1  &  RRab  \\
scl-CEMV032   &      V336	  &   ---   &  00 58 34.15  &  -33 51 25.4  &    0.5630753   &  20.574  & 20.215  & 19.692  &  1.101  &  0.980  &  0.637  &  0  &  1  &  RRab  \\
scl-CEMV033   &      V431	  &   ---   &  00 58 35.20  &  -33 46 12.2  &    0.3609695   &  20.382  & 20.138  & 19.679  &  0.649  &  0.522  &  0.259  &  0  &  1  &  RRc   \\
scl-CEMV034   &      V303	  &   ---   &  00 58 36.35  &  -33 41 37.3  &    0.3495360   &  20.442  & 20.201  & 19.803  &  0.612  &  0.517  &  0.325  &  0  &  1  &  RRc   \\
scl-CEMV035   &      V250	  &   ---   &  00 58 43.74  &  -33 42 06.1  &    0.6803114   &  20.415  & 20.074  & 19.474  &  1.108  &  0.852  &  0.577  &  0  &  1  &  RRab  \\
scl-CEMV036   &      ---	  &   ---   &  00 58 43.85  &  -33 21 58.1  &    0.337761    &  20.279  & 20.111  & 19.754  &  0.594  &  0.530  &  0.269  &  0  &  1  &  RRc   \\
scl-CEMV037   &      ---	  &   ---   &  00 58 44.07  &  -33 40 26.2  &    0.6682190   &  20.344  & 20.015  & 19.432  &  1.251  &  1.060  &  0.714  &  0  &  1  &  RRab  \\
scl-CEMV038   &      V246	  &   ---   &  00 58 44.60  &  -33 57 46.3  &    0.6700931   &  20.357  & 20.043  & 19.547  &  1.163  &  0.961  &  0.608  &  0  &  1  &  RRab  \\
scl-CEMV039   &      V556	  &   ---   &  00 58 45.04  &  -33 42 03.1  &    0.525160    &  20.505  & 20.215  & 19.900  &  1.167  &  1.029  &  0.460  &  3  &  0  &  RRab  \\
scl-CEMV040   &      ---	  &   ---   &  00 58 45.51  &  -33 38 32.8  &    0.3107837   &  20.469  & 20.239  & 19.848  &  0.698  &  0.533  &  0.325  &  3  &  0  &  RRc   \\
scl-CEMV042   &      ---	  &   ---   &  00 58 47.73  &  -33 34 01.5  &    0.3027722   &  20.412  & 20.194  & 19.839  &  0.689  &  0.568  &  0.343  &  0  &  1  &  RRc   \\
scl-CEMV043   &      ---	  &   ---   &  00 58 48.10  &  -33 55 48.6  &    0.2857029:  &  20.403  & 20.216  & 19.866  &  0.259  &  0.242  &  0.192  &  2  &  0  &  RRc   \\
scl-CEMV044   &      V485	  &   ---   &  00 58 48.46  &  -33 57 55.5  &    0.3550839   &  20.487  & 20.240  & 19.831  &  0.555  &  0.448  &  0.211  &  0  &  1  &  RRc   \\
scl-CEMV045   &      ---	  &   ---   &  00 58 49.27  &  -33 39 58.7  &    0.638994    &  20.447  & 20.122  & 19.630  &  0.637  &  0.618  &  0.424  &  2  &  0  &  RRab  \\
scl-CEMV046   &      V135	  &   ---   &  00 58 49.35  &  -33 47 17.0  &    0.5100380   &  20.466  & 20.186  & 19.786  &  1.494  &  1.262  &  0.723  &  0  &  1  &  RRab  \\
scl-CEMV047   &      V534	  &   ---   &  00 58 50.28  &  -33 30 31.4  &    0.4927699   &  20.551  & 20.271  & 19.790  &  1.320  &  1.030  &  0.715  &  2  &  0  &  RRab  \\
scl-CEMV049   &      V496	  &   ---   &  00 58 50.83  &  -33 51 21.2  &    0.3651721   &  20.168  & 19.972  & 19.585  &  0.385  &  0.313  &  0.255  &  3  &  0  &  RRc   \\
scl-CEMV050   &      V460	  &   ---   &  00 58 51.56  &  -33 38 06.7  &    0.3630319   &  20.428  & 20.175  & 19.762  &  0.529  &  0.442  &  0.169  &  2  &  0  &  RRd   \\
\hline
\end{tabular}
\begin{tablenotes}
\item $^{a}$ Q1 is a parameter associated with quality of the light curve: 0-good, 1-poor quality, 2-noisy, 3-outliers, 4-blended, 5-Blazhko effect.
\item $^{b}$ Q2 is a parameter related with the coverage of the light curve: 0-not well sampled and/or too noisy (\textit{full} RRLs sample), 
      1-well sampled and not too noisy (\textit{clean} RRLs sample). In the case of the 16 outliers, they are pointed out by an ``X''.
\item Full version are available as Supporting Information with the online 
version of the paper.
\end{tablenotes}
\end{scriptsize}
\end{table*}

Fig.~\ref{fig:cmd} shows that a few RRLs stars appear far from their expected 
location, presumably due to the poor coverage of the LC causing erroneous mean 
magnitudes in the different photometric bands. To avoid these outliers, following 
the procedure of \citetalias{MartinezVazquez2015} we selected those (520) with mean
\vv~ magnitude within 2.5$\sigma$ from the average of the population (20.13 mag): 
these define the \textit{full} sample that will be adopted in the analysis 
throughout the paper. On the other hand, a more restrictive
selection was performed based on the quality of the phase coverage of the
photometry over the entire pulsation cycle based on visual inspection of the
individual light curves and the period-magnitude diagram, resulting in the sample of 
290 RRL stars that we defined as the \textit{clean} sample (tinted background in 
Fig.~\ref{fig:rrl_lcv}).
In summary, we have identified 276 (167) RRab + 195 (123) RRc + 49 (0) candidate RRd 
variables in the \textit{full} (\textit{clean}) sample. 

	\subsection{Periods and Amplitudes}\label{sec:bailey}

\begin{figure}
	\includegraphics[scale=0.5]{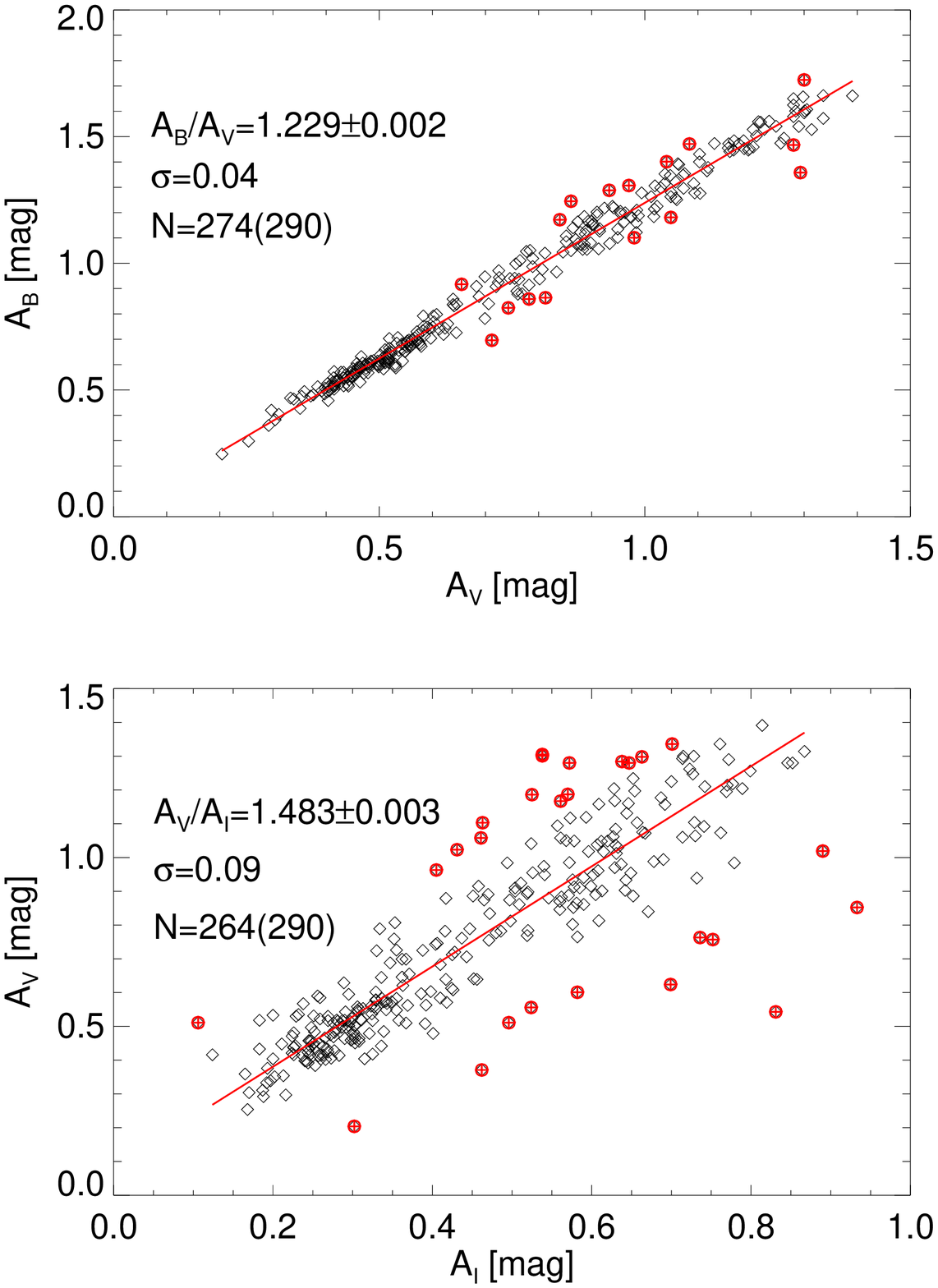}
	\caption{\bb-amplitude versus \vv-amplitude diagram (top) and 
	\vv-amplitude versus \ii-amplitude diagram (bottom). The slopes obtained 
	here are in good agreement with those predicted for the RRLs of the 
	Galactic GCs \citep{DiCriscienzo2011}. In both cases, the linear fit was 
	performed through the \textit{clean} sample of RRLs in Sculptor, applying 
	least squares fit and a 3$\sigma$ clipping.}
\label{fig:amplitude}
\end{figure}

Fig.~\ref{fig:amplitude} presents a comparison between the amplitudes in the
\bb~ vs \vv~ (upper panel) and \vv~ vs \ii~ bands (lower panel) for the {\itshape
clean}  sample of RRLs. We used them to perform a linear fit. The red symbols
show the outliers rejected with a 3-$\sigma$ clipping selection and not used in
the linear fit.  The derived values for the amplitude ratios, given by the slopes
of the red lines,  are 1.229$\pm$0.002 (A$_{\bb}$/A$_{\vv}$) and 1.483$\pm$0.003
(A$_{\vv}$/A$_{\ii}$). These values are in good agreement both with theoretical
predictions \citep{Bono1997a} and with observations of RRLs in Galactic globular
clusters \citep{DiCriscienzo2011}. This supports the accuracy of our derived
pulsation properties for Sculptor's RRL stars.

\begin{figure}
\includegraphics[scale=0.5]{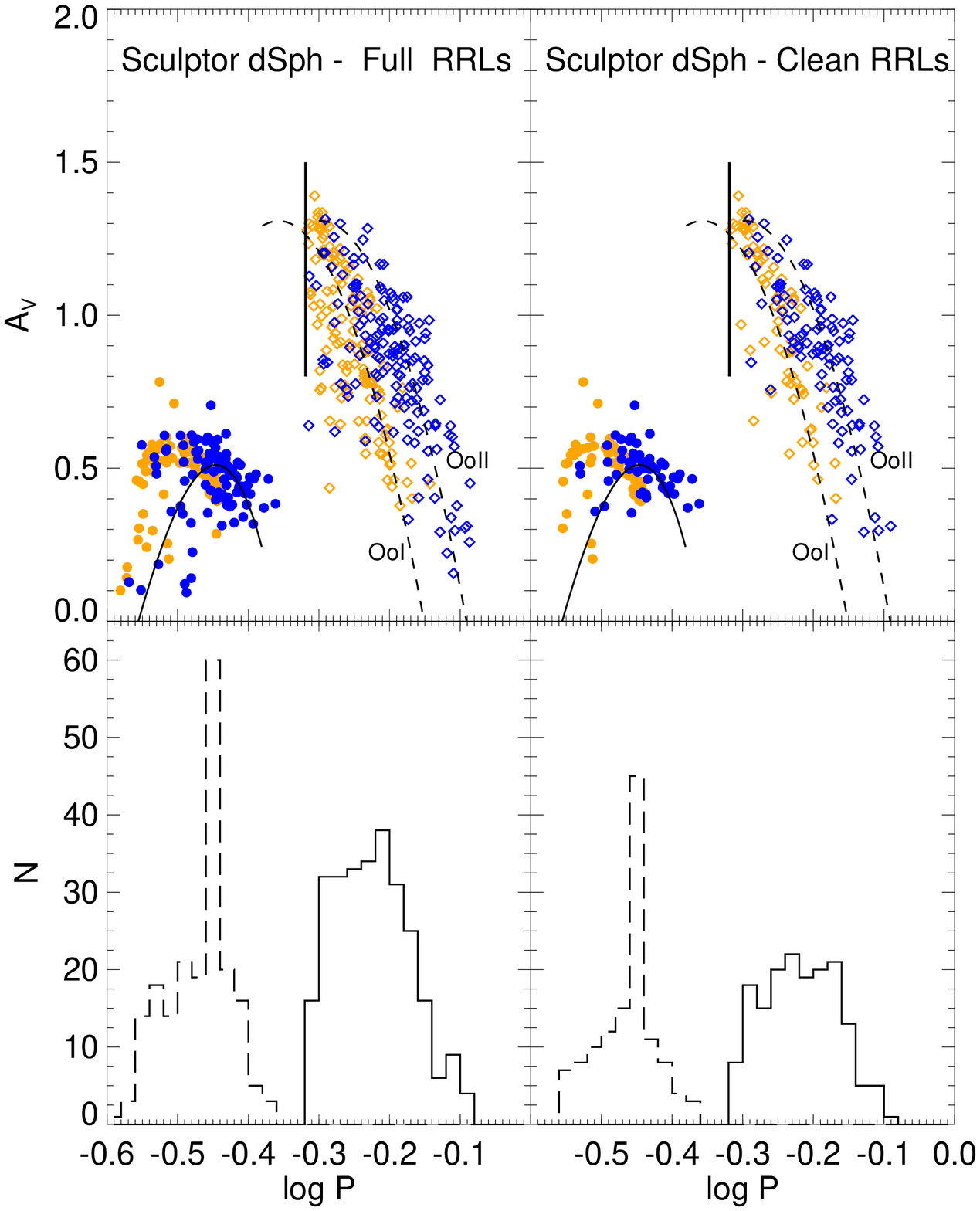}
\caption{Top. Period-amplitude or Bailey diagrams for the \textit{full} 
(\textit{left-hand panels}) and for the \textit{clean} (\textit{right-hand panels}) 
samples of RR Lyrae stars in Sculptor. Open diamonds and filled circles represent
 RRab and RRc, respectively. The dashed lines are the relations for RRab 
 stars in OoI and OoII clusters obtained by \citet{Cacciari2005} for the M3 RRab
 variables. The solid curve is derived from the M22 (OoII cluster) RRc variable
 study by \citet{Kunder2013}. Black vertical lines mark the HASP limit defined by 
 \citet{Fiorentino2015a}. Bottom. Period histograms for those RR Lyrae stars shown 
 in the top panels. RRab and RRc are shown as histograms with solid and dashed lines, 
 respectively. For the sake of clarity, RRd stars are not plotted as their
 periods are less certain.}
\label{fig:bailey}
\end{figure}

In Fig. \ref{fig:bailey}, we present the Bailey (period-amplitude) diagrams (top
panels) and the period distributions (bottom panels) for the \textit{full} (left
panels) and the \textit{clean} (right panels) samples of RRL stars. 
Galactic globular clusters (GGCs) with RRLs can be
classified into two groups \citep{Oosterhoff1939}, according to the mean period of
their RRab stars (Oo-I: 0.55 d, Oo-II: 0.64 d) and RRc stars (Oo-I: 0.32 d, Oo-II: 0.37 d).  
In the case of Sculptor, we find that the mean periods of the RRab and RRc stars are: 
$\langle P_{ab}\rangle$=0.602$\pm$0.004 d ($\sigma$=0.08) and 
$\langle P_{c}\rangle$=0.340$\pm$0.003 d ($\sigma$=0.04) for the \textit{full} sample, and 
$\langle P_{ab}\rangle$=0.609$\pm$0.006 d ($\sigma$=0.07) and 
$\langle P_{c}\rangle$=0.345$\pm$0.003 d ($\sigma$=0.03) for the \textit{clean} sample, 
thus placing Sculptor squarely in the so-called Oosterhoff gap \citep[see Fig.~5 from][]{Catelan2009}.  
Therefore, on the basis of the mean period of the RRab (and RRc) stars, Sculptor could be 
classified as an Oo-intermediate system, as is normal among Local Group dwarf galaxies
\citep{Kuehn2008, Bernard2009, Bernard2010, Garofalo2013, Stetson2014,
Cusano2015, Ordonez2016}.

Another tool used to classify GGCs in Oo-type is the Bailey diagram.
Oo-I and Oo-II clusters commonly follow the two distinct curves shown 
in Fig.~\ref{fig:bailey} (top panels) and defined in \citet{Cacciari2005}. 
The RRLs in Sculptor present a significantly wider distribution than either 
Oo-class. In fact, for each given amplitude stars cover a large period range 
around both Oo-lines, with a minor fraction populating the intermediate region. 
It is clear that an Oo-intermediate classification does not necessarily 
mean that stars are predominantly {\it between\/} the Oo-I and Oo-II loci; 
the distribution also extends to and beyond those loci. Therefore, a classification 
based only on the mean periods for RRab-type (and/or RRc-type) stars is clearly 
insufficient to characterize the range of properties among the RRL stars in Sculptor.  

As we demonstrated in \citetalias{MartinezVazquez2015}, the RRL stars show a
spread in \vv~magnitude of $\sim$0.35 mag, significantly larger than the typical
uncertainties in the mean magnitude ($\sigma$=0.03 mag), and larger than the spread
expected from the simple ageing of a mono-metallic population. We also show that the 
vast majority of the RRLs span a significant metallicity spread of $\sim$ 0.8 dex, 
bracketed between $\sim$--2.3 and $\sim$--1.5 dex. When splitting the sample of RRLs 
at $\langle$\vv$_{RRL}\rangle$=20.155 mag, thus defining a {\itshape bright} (Bt) and 
a {\itshape faint} (Ft) sample, the spread in metallicity is reflected in the two groups. 
In \citet[][\citetalias{MartinezVazquez2016}]{MartinezVazquez2016}, we show that 
the Bt sample is, on average, more metal-poor ($\langle$[Fe/H]$_{Bt}\rangle$ = --2.03) 
than the Ft one ($\langle$[Fe/H]$_{Ft}\rangle$ = --1.74).

The Bailey diagram (Fig.~\ref{fig:bailey}) shows the Bt and Ft samples
with blue and orange symbols, respectively. Interestingly, stars selected in the CMD 
are clearly separated in the Bailey diagram as well: Bt, metal-poor RRL stars are closer 
to the Oo-II sequence, while Ft, more metal-rich RRL stars follow a distribution similar to
an Oo-I system. This is also reflected in the mean periods of the Bt and Ft groups, which 
are similar to those defining the OoII and OoI systems, respectively (see 
Table~\ref{tab:mean_period}).
This supports the conclusion that in complicated systems such as Sculptor,
characterized by an important chemical evolution at an early epoch, the
Oosterhoff classification must be treated cautiously, and that the mean period
alone does not provide a full characterisation of the target stellar system.

\begin{table}
\begin{scriptsize}
\centering
\caption{Mean period of the Bt and Ft groups for both \textit{full} and \textit{clean} RRL samples.} 
\label{tab:mean_period}
  \begin{tabular}{ccc}
 \hline
 & \textbf{$\langle P_{ab}\rangle$} & \textbf{$\langle P_{c}\rangle$} \\
\hline 
\textbf{FAINT (Ft)} & & \\
\hline
\textit{full} & 0.560$\pm$0.004 d ($\sigma$=0.05) & 0.325$\pm$0.007 d ($\sigma$=0.03) \\
\textit{clean} & 0.560$\pm$0.006 d ($\sigma$=0.05) & 0.332$\pm$0.003 d ($\sigma$=0.03) \\
\hline
\textbf{BRIGHT (Bt)} & & \\
\hline
\textit{full} & 0.639$\pm$0.006 d ($\sigma$=0.08) & 0.356$\pm$0.006 d ($\sigma$=0.03) \\
\textit{clean} & 0.647$\pm$0.007 d ($\sigma$=0.07) & 0.362$\pm$0.004 d ($\sigma$=0.03) \\
\hline
\end{tabular}
\end{scriptsize}
\end{table}

Fig. \ref{fig:bailey} also provides constraints on the old galaxy stellar population:

\textit{(i)} The fraction of RRc stars in the \textit{full} sample,
$f_c$=$\frac{N_{c}}{N_{ab}+N_{c}}$=0.41 ($f_c$=0.42 for the \textit{clean} sample of RRLs) 
in agreement with \citealt{Kaluzny1995}, is relatively higher than in the rest of the other dSph
galaxies (see Table~6 of \citealt{Stetson2014}): it is almost
twice the $f_c$ obtained in other dSphs. This may be linked to the HB 
morphology, which has a strong blue component in Sculptor, which may in turn
be linked to the details of its early star-formation history.

\textit{(ii)} The shortest period found for the RRab stars is 0.48230141 days
(with A$_{\vv}$=1.280 mag, star: scl-CEMV397). The lack of High Amplitude Short
Period fundamental RRL stars \citep[HASP; A$_{\vv} \gtrsim $ 0.75 mag, P $\leq$
0.48 (log P $\leq$ -0.32)][]{Fiorentino2015a} suggests that Sculptor does not host 
a significant metal-rich ([Fe/H]$\gtrsim$--1.5) old stellar population, as has been 
confirmed in our analysis of \citetalias{MartinezVazquez2015}.

	\subsection{Distance to Sculptor from the RRL stars}\label{sec:distance}

In \citetalias{MartinezVazquez2015} we estimated the true distance modulus from
the \textit{full} sample as well as the \textit{clean} sample of RRLs. The distance
estimates to Sculptor were derived using the three different subsamples (RRab,
RRc, and RRab+RRc fundamentalized). The use of three photometric bands and
two period-Wesenheit relations (PWR) in \vv, \bmv, and \vv, \bmi~ (these are
reddening-free and are only minimally affected by the metal content,
\citealt{Marconi2015}) allowed us to provide a very accurate distance. Table~3 in
\citetalias{MartinezVazquez2015} summarizes the individual distance moduli
obtained applying different calibrations\footnote{ It is worth mentioning 
that Ft and Bt samples produce similar distance moduli.}. The errors on individual distance
moduli are never larger than 0.02$_{sys}$ and 0.09$_{ran}$. Finally, we adopted
a mean true distance modulus ($\mu$) of 19.62 mag with $\sigma$=0.04 (see \S~4
of \citetalias{MartinezVazquez2015} for more details). This estimate is in
good agreement with values based on other reliable indicators
\citep{Rizzi2002,Pietrzynski2008}. 

	\subsection{Comparison with the Kaluzny catalogue}\label{sec:kaluzny_comparation}

\citet{Kaluzny1995} published a list of 226 RRL stars covering the central
15$\arcmin\times$15$\arcmin$ of Sculptor. We matched the two catalogues in order to check
the consistency of the derived pulsational properties. 216 of Kaluzny's RRL stars
are included in our sample.  It is worth mentioning that the work of
\citet{Kaluzny1995} includes 226 sources, but we realized that: {\itshape
i)} 5 of these stars were duplicated (K1926=K406, K2558=K2058, K2559=K2059,
K3345=K1439, K4233=K2423); {\itshape ii)} one (K403) is not variable in our
photometry; {\itshape iii)} the variability of K5081 is not certain based on our
data; {\itshape iv)} one more star is possibly misclassified, and we catalogue
it as probable eclipsing binary (K3710, \citealt{Clementini2005}); {\itshape v)} for two 
of the Kaluzny's stars (K737, K4780) we are not able to derive a reliable period. 
The 3 AC (K3302-V25, K734-V119, and K5689) and the 2 LPV (K274 and K687) analysed 
in \cite{Kaluzny1995} were also found in our catalogue so, in total, we matched 224 
variable stars (216 RRL + 3 AC + 2 LPV + 1 eclipsing binary + 2 probable variable stars).

The individual and global properties they found are in good agreement with the
values we redetermined. In particular, 90 percent of the stars matched have the same
periods in both  catalogues, within 0.001 days. The global properties are also 
in good agreement, despite the fact that the two surveys cover a different fraction
of Sculptor's main body. \cite{Kaluzny1995} found that the mean period of
the fundamental pulsators and the frequency of first overtones are,
respectively, $<P_{ab}>$=0.585 d, $<P_{c}>$=0.338 d,  and $f_c$=0.40. In this
work, the mean periods of the 289 RRab and 197 RRc are
$<P_{ab}>$=0.602$\pm$0.004 d ($\sigma$=0.08), and $<P_{c}>$=0.340$\pm$0.003 d
($\sigma$=0.04), respectively. The fraction of RRc variables is equal to 
$f_c$=0.41 and becomes 0.46 if we also include the RRd, i.e., $f_{cd}$=0.46.
It has to be stressed
that roughly 2/3 of the OGLE images are included in our sample. The few outliers
with discrepant periods in the two studies can be ascribed to an aliasing problem, 
which is more likely solved by our larger database. Nevertheless, the overall excellent 
agreement is further independent proof of the quality of the OGLE data and observing 
strategy, even with a limited number of phase points as in this case.
 
\section{Anomalous Cepheid stars} \label{sec:acep}

We confirm the existence of 4 Anomalous Cepheids (ACs) in Sculptor, as
previously found by \citet{Smith1986} and \citet{Kaluzny1995}. In
Table~\ref{tab:acep} we summarize the properties of these stars. Two of them
(scl-CEMV284 and scl-CEMV388) were discovered by \cite{Baade1939} and analysed
for the first time by \cite{Swope1968}, obtaining periods of 0.926 and 1.346
days, respectively. Another (scl-CEMV160) was discovered by
\cite{Thackeray1950}, who obtained a period of 1.15 days. All three of these
variables were included in the \cite{vanAgt1978} catalogue, with similar
periods. The fourth AC, scl-CEMV447, was discovered and classified as such by
\cite{Kaluzny1995}, with a period of 0.85541 days. These previous determinations
of the periods for the four ACs are in good  agreement with those presented here
(see the sixth column in Table~\ref{tab:acep}).

\begin{table*}
\begin{scriptsize}
\centering
\caption{Parameters of the AC stars in Sculptor dSph.} 
\label{tab:acep}
  \begin{tabular}{ccccccccccccccc} 
 \hline
CEMV+2016 & Original & Alternative & RA & DEC & Period & <\bb> & <\vv> & <\ii> & A$_{\bb}$ & A$_{\vv}$ & A$_{\ii}$ & Q1 & Q2 & Type \\
name & name & name & (J2000) & (J2000) & (current) & & & & & & & & & \\ 
\hline 
scl-CEMV160   &      V119  	&	  K734   &  00 59 33.93  &  -33 39 31.0  &    1.1577836   &  19.221  & 18.880  & 18.284  &  0.703  &  0.647  &  0.343  &  0  &  1  & F \\
scl-CEMV284   &      V25	  &	  ---    &  00 59 58.53  &  -33 33 42.2  &    0.9255408   &  19.095  & 18.812  & 18.379  &  1.600  &  1.195  &  0.737  &  0  &  1  & F \\
scl-CEMV388   &      V26	  &	  K3302  &  01 00 16.64  &  -33 47 57.4  &    1.3460368   &  19.007  & 18.623  & 18.097  &  1.045  &  0.863  &  0.536  &  0  &  1  & F \\
scl-CEMV447   &      K5689	&	  ---	   &  01 00 30.32  &  -33 41 42.5  &    0.8554217   &  19.472  & 19.133  & 18.617  &  0.941  &  0.715  &  0.381  &  2  &  0  & F \\
\hline
\end{tabular}
\begin{tablenotes}
\item See the caption of Table~\ref{tab:rrl} for a description of the Q1 and Q2 parameters.
\end{tablenotes}
\end{scriptsize}
\end{table*}

ACs can form through two different channels \citep{Bono1997b,Cassisi2013}. They
can be the progeny of  coalesced binary stars, thus evolved blue straggler
stars (BSS) tracing the old population
\citep{Renzini1977,Hirshfeld1980,Sills2009}. Alternatively, they can  be a life stage
of metal-poor (Z<0.0006, \citealt{Fiorentino2006}), single stars with
mass between $\sim$1.2 and $\sim$2.2 M$_{\sun}$ and age between 1 and 6 Gyr
\citep{Demarque1975,Norris1975,Castellani1995,Caputo1999}. Given the
predominantly old stellar population of Sculptor \citep{deBoer2012}, it is
unlikely that ACs proceed from such a young, elusive population (if any), in
agreement also with the analysis of the BSS population \citep{Mapelli2009} 
which excludes the occurrence of such young stars in Sculptor.

\begin{figure}
	\includegraphics[scale=0.5]{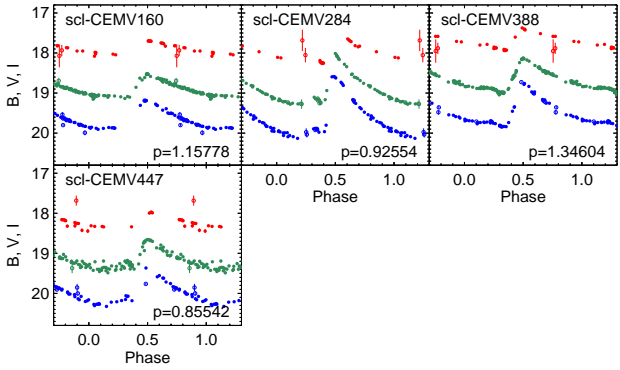}
	\vspace{-7cm}
	\caption{Sample of light curves of the AC stars in the \bb~ (blue), \vv~ (green) and
	 \ii~ (red) bands, phased with the period in days given in the lower right-hand corner
	  of each panel. The name of the variable is given in the left-hand corner of each panel. 
		Open symbols show the bad data points, i.e., with discrepancies larger than 3$\sigma$ above 
		the standard error of a given star; these were not used in the calculation of the period 
		and mean magnitudes. For clarity, the \bb~ and \ii~ light curves have been shifted 
		by 0.4 mag down- and upward, respectively.}
	\label{fig:acep_lcv}
\end{figure}

The classification of the pulsation mode of ACs is not trivial and cannot be
easily determined from the morphology of the light curves
(Fig.~\ref{fig:acep_lcv}) or from the period-amplitude diagram alone. Theoretical
predictions indicate that for the same luminosity and colour, first-overtone
(FO)  pulsators are less massive than fundamental (F) pulsators. Therefore,
determining  the correct pulsational mode is important 
for obtaining a reliable mass estimate. To distinguish between F and FO pulsators 
we follow two approaches. First, the different pulsation modes follow different period-luminosity 
relations -- which are also different from those of classical and type II cepheids. 
Their location in the PL diagram can therefore be used to constrain their pulsation mode 
\citep[see, e.g.,][]{Bernard2013}. We find that all four ACs in Sculptor fall squarely on 
the sequence corresponding to the fundamental mode ACs. The second approach has been 
presented in \citet{Fiorentino2012b}, and is based on the method described in 
\citet{Marconi2004}. It is known that ACs obey 
well-defined mass-dependent period-luminosity-amplitude (MPLA) and period-luminosity-colour
(MPLC) relations \citep{Marconi2004}:
{\small 
\begin{equation}
\log \dfrac{M_{MPLA, F}}{M_{\sun}} = (0.01-0.188 \cdot A_{\vv}-\logp-0.41 \cdot M_{\vv})/0.77
\end{equation}

\begin{equation}
\log \dfrac{M_{MPLC, F}}{M_{\sun}} = -(M_{\vv}+1.56+2.85 \cdot \logp-3.51 \cdot (M_{\bb}-M_{\vv}))/1.88
\end{equation}

\begin{equation}
\log \dfrac{M_{MPLC, FO}}{M_{\sun}} = -(M_{\vv}+1.92+2.90 \cdot \logp-3.43 \cdot (M_{\bb}-M_{\vv}))/1.82
\end{equation}
}

However, the MPLA relation is only valid for F pulsators, whereas MPLC relations 
exist for both pulsation modes. 

In order to assign a pulsation mode and a mass to each AC, we proceed as follows.
We first estimate the mass using both the MPLA and MPLC relations for F pulsators. 
Then, when these two values agree with each other within 2-$\sigma$,
we classify the star as F mode and take the mean mass as the true value.
If instead the two mass estimates are not consistent, we assume that the star is a FO
pulsator and we use the corresponding MPLC relation for the mass estimation.
This method confirms independently that all ACs are F pulsators (see Table~\ref{tab:mass_acep}).

\begin{table*}
\begin{scriptsize}
\centering
\caption{Parameters of the AC stars in Sculptor dSph.} 
\label{tab:mass_acep}
\begin{tabular}{cccccc} 
\hline
CEMV+2016 & M$_{MPLA, FU}$ & M$_{MPLC, FU}$ & M$_{MPLC, FO}$ & MODE$_{mass}$ & $<$M$>^{*}$ \\
name & M$_{\sun}$ & M$_{\sun}$ & M$_{\sun}$ & & M$_{\sun}$ \\ 
\hline 
scl-CEMV160 & 1.56$\pm$0.18 & 1.23$\pm$0.06 & 0.77$\pm$0.04 & F & 1.40$\pm$0.19 \\
scl-CEMV284 & 1.68$\pm$0.19 & 1.50$\pm$0.07 & 0.94$\pm$0.04 & F & 1.59$\pm$0.21 \\
scl-CEMV388 & 1.57$\pm$0.18 & 1.65$\pm$0.08 & 1.02$\pm$0.05 & F & 1.61$\pm$0.20 \\
scl-CEMV447 & 1.64$\pm$0.19 & 1.45$\pm$0.07 & 0.90$\pm$0.04 & F & 1.54$\pm$0.20 \\
\hline
\end{tabular}
\begin{tablenotes}
\item $^{*}$The final mean value is the mean of the fundamental estimates.
\end{tablenotes}
\end{scriptsize}
\end{table*}

\citet{Smith1986} provided a mass value for two Sculptor ACs: scl-CEMV284
(2.0$^{+1.4}_{-0.8}$) M$\sun$ and scl-CEMV388 (0.6$^{+0.4}_{-0.2}$M$\sun$) from
a linear pulsation theory assuming that they pulsate in the F mode with a
pulsational constant of $Q$=0.0034 \citep{Wallerstein1984}. The first value is
in agreement within the errors with our estimate, but the second one is
different at the 3-$\sigma$ level. Nevertheless, as \citet{Smith1986}  noted,
both their mass estimates have substantial uncertainties due to their 
photometry, and our improved data make us more confident in our determinations.

\begin{figure*}
	\includegraphics[scale=0.6]{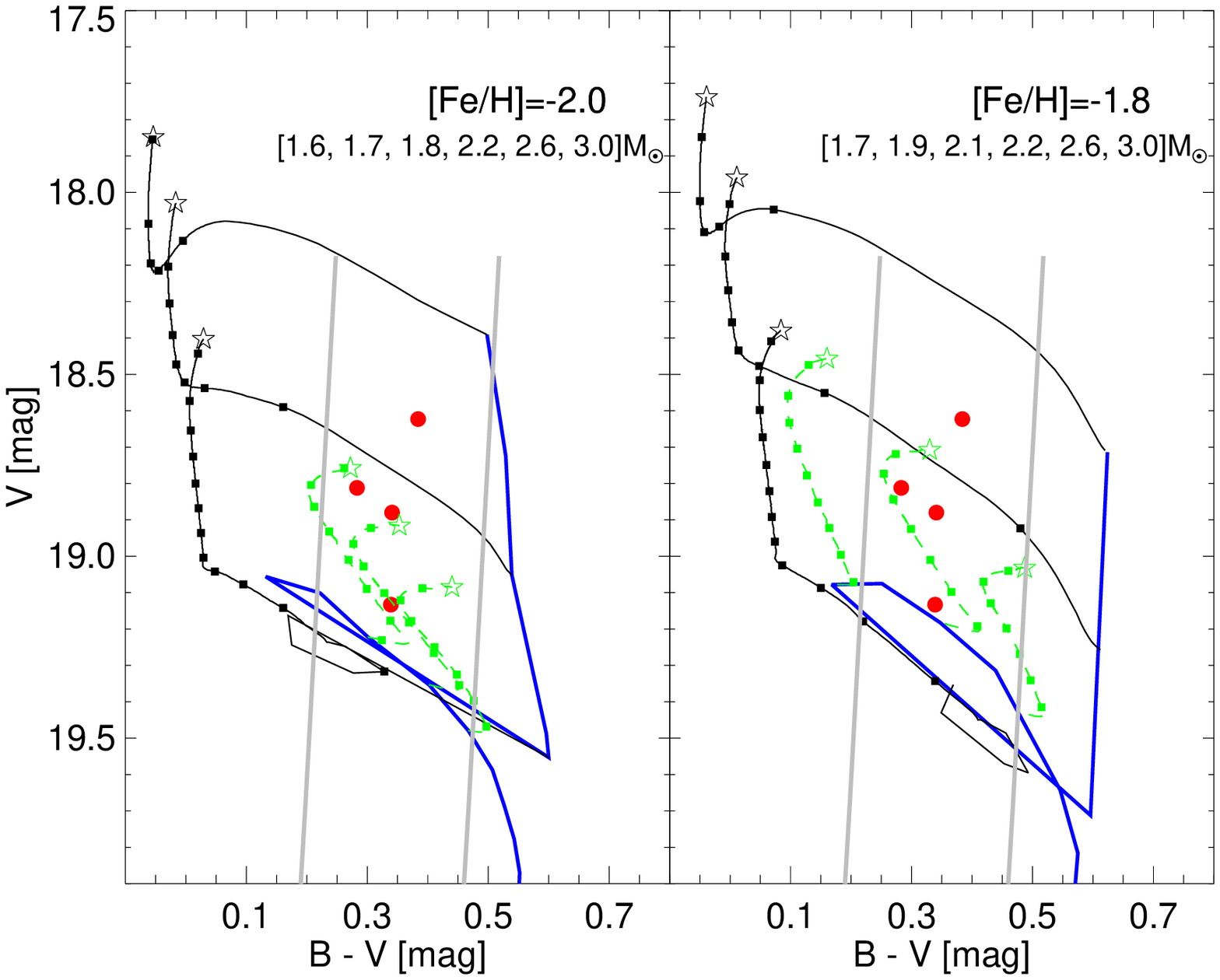}
	\caption{Optical (\bmv, \vv) CMDs of Sculptor, zoom-in on AC stars, where 
	theoretical predictions from BaSTI library \citep{Pietrinferni2004} 
	are over-plotted. Solid grey lines represent the theoretical 
	instability strip \citep{Fiorentino2006}. The red dots indicate the 4 ACs found
	in Sculptor. The blue lines show the zero-age helium-burning (ZAHeB) 
	loci for stars with masses in the range between $\sim$1.0 and 3.0 
	M$_{\sun}$. Evolutionary tracks for AC stars for the labelled masses
  are shown for different metallicities: [Fe/H]=-2.0 (left), [Fe/H]=-1.8(right). 
	Green and black lines	indicate models that ignite helium in the core in 
	degenerate and non-degenerate conditions, respectively. These stellar 
	tracks represent the path from the ZAHeB to the central helium 
	exhaustion (10 percent of the initial abundance), indicated by an open star 
	symbol, at different stellar masses. For the theoretical analysis we 
	used scaled-solar evolutionary models, with a fixed $\Delta 
	Y / \Delta Z$=1.4 and a primordial $Y$=0.245, assumed a distance modulus 
	of $\mu$=19.62 mag (\citetalias{MartinezVazquez2015}) and a reddening 
	of 0.018 mag \citep{McConnachie2012}.}
	\label{fig:ac_tracks}
\end{figure*}

Fig~\ref{fig:ac_tracks} is a zoom-in of the CMD on the region of AC stars  that
shows a comparison of theoretical evolutionary tracks for different  masses (1.5
-- 3.0 M$_{\sun}$) and [Fe/H]~(--2.0 and --1.8). The adopted stellar models have
been retrieved from the BaSTI library \citep{Pietrinferni2004} or---when not 
available in the database---have been computed for this specific project using a 
theoretical framework fully consistent with the BaSTI assumptions. In more detail,  
the models shown in Fig.~\ref{fig:ac_tracks} are based on a scaled-solar heavy element
mixture, assume a canonical (no overshooting) physical  framework, and a
mass loss efficiency on the RGB with the free parameter $\eta$  entering in the
\citet{Reimers1975} law set to 0.4. The models were shifted  using the distance
modulus $\mu$=19.62 mag (\citetalias{MartinezVazquez2015})  and foreground
reddening E(\bmv)=0.018 mag \citep{Pietrzynski2008}, and assuming  the standard
extinction law from \cite{Cardelli1989}. The blue lines show the zero-age 
helium-burning (ZAHeB) loci for stars with masses from $\sim$1 to 3 M$_{\sun}$. 
Green dashed and black solid lines correspond to the core He-burning
evolutionary tracks for stellar models igniting triple-$\alpha$ nuclear
reactions at the tip of the RGB under  conditions of partial electron degeneracy
(initial mass lower than $\sim$2.0M$_{\sun}$)  and no-electron degeneracy,
respectively (see \citealt{Cassisi2013} for a  detailed review on
this topic). The termination of the core He-burning stage (central  He abundance
equal to 10 percent of the initial value) is marked as green and black stars, 
respectively. The gray lines correspond to the boundaries of the instability
strip \citep{Fiorentino2006}.

The left panel of Fig.~\ref{fig:ac_tracks} shows that a metallicity of [Fe/H]=--2.0
provides a good match between the location of three of the observed stars and
the tracks with mass value in the range 1.60--1.80 M$_{\sun}$. These masses are in
good agreement, within the errors, with the mean mass estimated above: 
$<M_{AC}>$=1.54M$_{\sun}$ ($\sigma_{sys}$=0.20 M$_{\sun}$,
$\sigma_{ran}$ = 0.08 M$_{\sun}$).  
However, the brightest AC is $\sim 0.1$ mag brighter than expected at this metallicity.
An increase in metallicity by 0.2 dex (right panel) provides a good fit of all four stars, 
but implies slightly larger stellar masses of $\sim$1.90 M$_{\sun}$ that is in agreement, 
for the adopted theoretical scenario, within 2-$\sigma$.

These two low metallicities are in agreement with previous estimates by \citet{Smith1986}, 
who gave a mean value close to [Fe/H]=--1.9 ($\sigma$ = 0.3 dex)\footnote{\cite{Smith1986} 
used the $\Delta$S parameter to measure the metallicities for the ACs through an adaptation 
of the $\Delta$S method for RRLs.} as the mean of three out the four ACs (scl-CEMV284, 
scl-CEMV388 and scl-CEMV160).

\section{SX Phoenicis stars}\label{sec:sxpho}

\begin{figure}
	\hspace{-1cm}
	\includegraphics[scale=0.5]{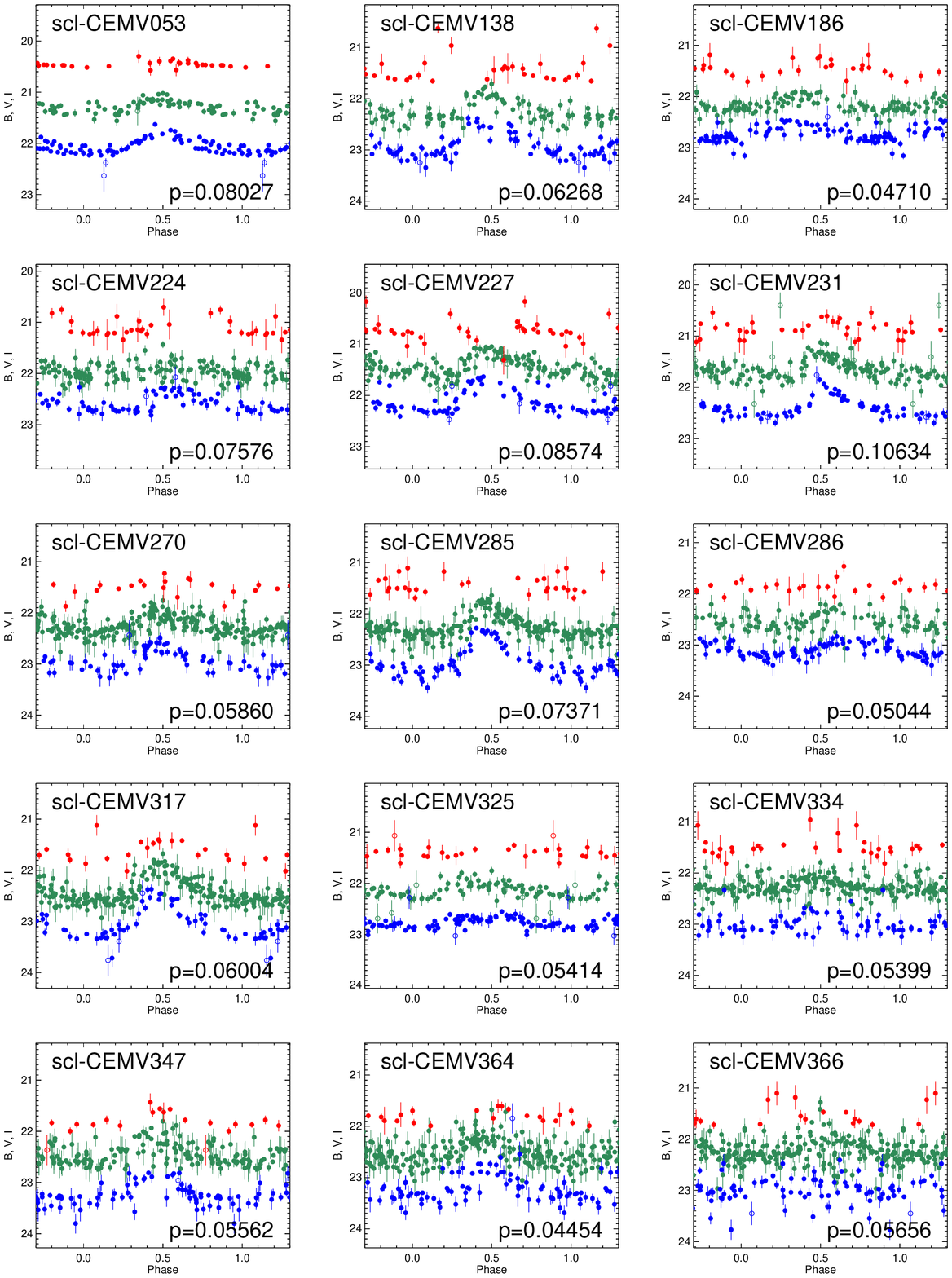}
	\caption{Sample of light curves of the SX~Phe stars in the \bb~ (blue), \vv~ (green) and \ii~ 
		(red) bands, phased with the period in days given in the lower right-hand 
		corner of each panel. The name of the variable is given in the left-hand 
		corner of each panel. Open symbols show the bad data points, i.e., with discrepancies 
		larger than 3$\sigma$ above the standard error of a given star; these were not used 
		in the calculation of the period and mean magnitudes. For clarity, the \bb~ and 
		\ii~ light curves have been shifted by 0.4 mag down- and upward, respectively. 
		Beige backgrounds mark those SX~Phe stars classified as within the \textit{clean} sample. 
		All light curves are available as Supporting Information with the online 
		version of the paper.}
	\label{fig:sxpho_lcv}
\end{figure}

Fig.~\ref{fig:sxpho_lcv} shows the light curves of the 23 variable stars
identified in the relatively faint part of the CMD of Sculptor, where  the IS
crosses the MS and the sub-giant branch for masses typically larger than 1 
M$_{\sun}$ (see Fig.~\ref{fig:cmd}). In this region, variable stars are 
characterized by short periods (from minutes to a few hours) and low amplitudes
(a few tenths of a magnitude). They typically present several pulsation modes
simultaneously, both radial and non-radial \citep{Santolamazza2001,Poretti2008}. 
These properties makes them elusive
targets in external galaxies, as the intrinsic faint brightness makes
them difficult to detect, given that the long exposure times required to have
good measurements conflict with their short periods. Commonly, these variable 
stars are classified as $\delta$~Scuti when they are population I (young and more
metal-rich stars), or SX Phoenicis (SX~Phe) when they are population II metal-poor
counterparts. The latter are typically observed in GCs, and they are associated
with the BSS that, leaving the MS, cross the IS {\itshape above} the main-sequence 
turnoff of the cluster population. SX~Phe stars are important because their
pulsational properties allow us to derive their distances \citep{McNamara2011}
and structural parameters such as the BSS mass \citep{Fiorentino2014,Fiorentino2015b}, 
which is a key ingredient in deriving the dynamical friction in globular clusters
\citep{Ferraro2012}.

In the case of dwarf galaxies the situation is more complicated. On the one
hand, dwarf galaxies that are composed of old populations only ($>$ 10 Gyr, or 
\textit{fast} systems according to the nomenclature introduced by
\citealt{Gallart2015}) appear similar to GCs: the colour of the main-sequence
turnoff (MSTO) stars is redder than the red edge of the IS. Therefore, only BSS
stars (bluer and brighter than the MSTO stars) can cross the IS at this
magnitude level when moving red-ward after central H exhaustion. On the other
hand, \textit{slow} dwarf galaxies  characterized by star formation at all
epochs can host a mix of different variable stars in this region of the CMD (see
the case of Carina, \citealt{Coppola2015} and the Large Magellanic Cloud,
\citealt{Poleski2010}). Given the lack of evidence for the  presence of a young 
population in Sculptor (see \S~\ref{sec:acep}), we assume here that these are 
the progeny of coalesced binaries stars, and therefore SX~Phe.

SX~Phe stars of different pulsational modes do not show a clear separation in the 
period-amplitude diagram. However, they do follow a period-luminosity relation
\citep{McNamara2000, Petersen1999, Fiorentino2014}.

Considering the above evidence and taking advantage of the models/relations
developed in \cite{Fiorentino2015b}, we obtain the theoretical
pulsation parameters for the 23 SX~Phe stars discovered for the first time in
Sculptor (see Table~\ref{tab:sxpho}). First, we used the predicted PL relations
in the \bb~ and \vv~ bands to perform a detailed mode identification, assuming
the distance modulus derived in \citetalias{MartinezVazquez2015} and reddening
from \citet{Pietrzynski2008}. Pulsation
modes can best be identified using the theoretical predictions in the \bb-band
PL planes (see Table~3 in \citealt{Fiorentino2015b}) because the \bb-band  is
the least noisy for this kind of star. We considered the relation for
metallicities Z=0.001 and Z=0.0001. The pulsation modes (F-fundamental-, FO-first
overtone-, SO-second overtone-, and TO-third overtone-) have been determined 
according to the proximity to a PL relation in the M$_{\bb}$-period plane.

\begin{table*}
\begin{scriptsize}
\centering
\caption{Parameters of the SX~Phe stars in Sculptor dSph.} 
\label{tab:sxpho}
  \begin{tabular}{ccccccccccccccc} 
 \hline
CEMV+2016 & Original & Alternative & RA & DEC & Period & <\bb> & <\vv> & <\ii> & A$_{\bb}$ & A$_{\vv}$ & A$_{\ii}$ & Q1 & Q2 \\
name & name & name & (J2000) & (J2000) & (current) & & & & & & & & & \\ 
\hline 
scl-CEMV053   &      ---  &	  ---	 &  00 58 52.18  &  -33 50 13.4  &    0.08027222  &  21.638  & 21.307  & 20.870  &  0.414  &  0.266  &  0.128  &  2  &  0   \\
scl-CEMV138   &      ---	&	  ---	 &  00 59 27.90  &  -33 49 35.4  &    0.06268482  &  22.433  & 22.206  & 21.683  &  0.683  &  0.535  &  1.098  &  2  &  0   \\
scl-CEMV186   &      ---	&	  ---	 &  00 59 38.98  &  -33 50 14.0  &    0.04709882  &  22.291  & 22.134  & 21.874  &  0.318  &  0.279  &  0.213  &  2  &  0   \\
scl-CEMV224   &      ---	&	  ---	 &  00 59 47.17  &  -33 38 16.3  &    0.07575704  &  22.138  & 21.932  & 21.425  &  0.437  &  0.318  &  0.493  &  2  &  0   \\
scl-CEMV227   &      ---  &	  ---	 &  00 59 49.29  &  -33 46 16.6  &    0.08574464  &  21.632  & 21.378  & 21.096  &  0.703  &  0.475  &  0.386  &  2  &  0   \\
scl-CEMV231   &      ---	&	  ---	 &  00 59 51.03  &  -33 41 05.4  &    0.10634337  &  21.967  & 21.624  & 21.208  &  0.511  &  0.511  &  0.286  &  2  &  0   \\
scl-CEMV270   &      ---	&	  ---	 &  00 59 56.29  &  -33 42 00.3  &    0.05859806  &  22.491  & 22.221  & 21.825  &  0.450  &  0.404  &  0.318  &  2  &  0   \\
scl-CEMV285   &      --- 	&	  ---	 &  00 59 58.67  &  -33 42 19.8  &    0.07370760  &  22.425  & 22.181  & 21.775  &  0.842  &  0.647  &  0.503  &  2  &  0   \\
scl-CEMV286   &      ---	&	  ---	 &  00 59 58.62  &  -33 50 09.7  &    0.05043523  &  22.689  & 22.516  & 22.257  &  0.273  &  0.320  &  0.126  &  2  &  0   \\
scl-CEMV317   &      ---  &	  ---	 &  01 00 03.08  &  -33 44 46.9  &    0.06004051  &  22.515  & 22.307  & 22.004  &  0.845  &  0.775  &  0.400  &  2  &  0   \\
scl-CEMV325   &      ---	&	  ---	 &  01 00 04.57  &  -33 52 15.4  &    0.05413676  &  22.363  & 22.126  & 21.793  &  0.209  &  0.234  &  0.174  &  2  &  0 	\\  
scl-CEMV334   &      ---	&	  ---	 &  01 00 05.75  &  -33 43 33.8  &    0.05398612  &  22.539  & 22.248  & 21.897  &  0.304  &  0.253  &  0.090  &  2  &  0   \\
scl-CEMV347   &      ---	&	  ---	 &  01 00 07.56  &  -33 39 51.0  &    0.05561832  &  22.726  & 22.381  & 22.159  &  0.631  &  0.582  &  0.482  &  2  &  0   \\
scl-CEMV364   &      ---	&	  ---	 &  01 00 11.00  &  -33 40 51.2  &    0.04454330  &  22.723  & 22.514  & 22.181  &  0.525  &  0.446  &  0.226  &  2  &  0   \\
scl-CEMV366   &      ---	&	  ---	 &  01 00 11.57  &  -33 41 27.5  &    0.05656062  &  22.534  & 22.184  & 21.773  &  0.082  &  0.326  &  0.449  &  2  &  0 	\\  
scl-CEMV383   &      ---	&	  ---	 &  01 00 15.43  &  -33 44 05.1  &    0.09018717  &  21.689  & 21.414  & 21.048  &  0.350  &  0.342  &  0.211  &  2  &  0   \\
scl-CEMV394   &      ---	&	  ---	 &  01 00 18.03  &  -33 47 48.5  &    0.08420236  &  22.168  & 21.926  & 21.596  &  0.724  &  0.515  &  0.398  &  2  &  0   \\
scl-CEMV410   &      ---	&	  ---	 &  01 00 20.06  &  -33 46 50.6  &    0.06548483  &  21.989  & 21.796  & 21.459  &  0.578  &  0.473  &  0.250  &  2  &  0   \\
scl-CEMV437   &      ---	&	  ---	 &  01 00 28.18  &  -33 50 11.3  &    0.06370454  &  22.434  & 22.260  & 21.845  &  0.592  &  0.386  &  0.618  &  2  &  0   \\
scl-CEMV500   &      ---	&	  ---	 &  01 00 40.84  &  -33 51 15.3  &    0.06184286  &  22.551  & 22.329  & 22.051  &  0.768  &  0.728  &  0.399  &  2  &  0   \\
scl-CEMV538   &      ---	&	  ---	 &  01 00 55.95  &  -33 31 40.1  &    0.05790988  &  22.464  & 22.173  & 21.783  &  0.285  &  0.236  &  0.075  &  2  &  0   \\
scl-CEMV540   &      ---	&	  ---	 &  01 00 56.73  &  -33 40 14.0  &    0.06031101  &  22.659  & 22.341  & 22.028  &  0.303  &  0.289  &  0.207  &  2  &  0   \\
scl-CEMV541   &      ---	&	  ---	 &  01 00 56.60  &  -33 43 29.8  &    0.06464694  &  22.424  & 22.242  & 21.854  &  0.507  &  0.369  &  0.439  &  2  &  0   \\
\hline
\end{tabular}
\begin{tablenotes}
\item See the caption of Table~\ref{tab:rrl} for a description of the Q1 and Q2 parameters.
\end{tablenotes}
\end{scriptsize}
\end{table*}

\begin{figure}
	\includegraphics[scale=0.6]{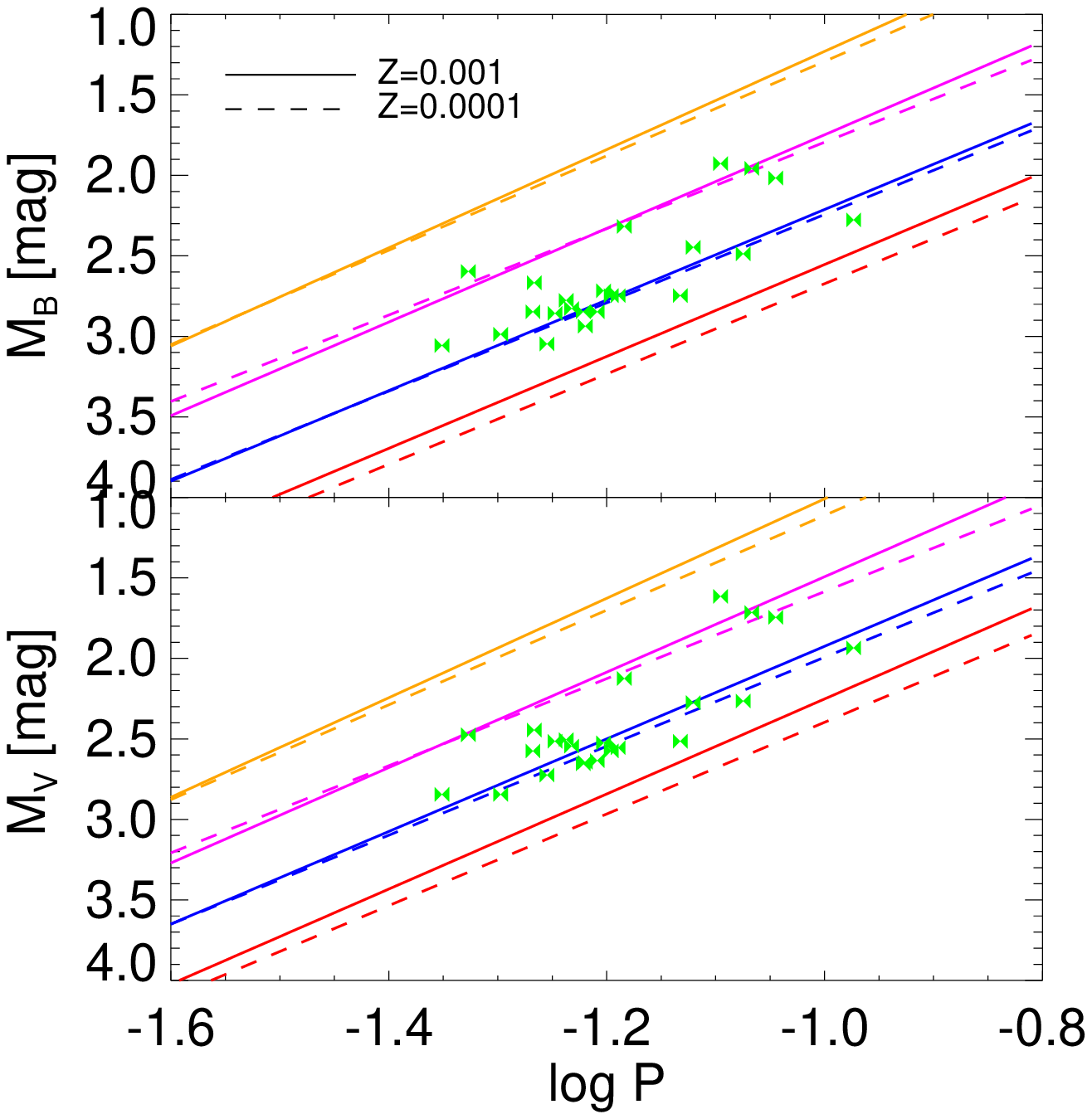}
	\caption{Absolute M$_{\bb}$ (top) and M$_{\vv}$ (bottom) period-luminosity 
	diagrams with theoretical PL relation of SX~Phe stars for Z=0.001 (solid lines) 
	and Z=0.0001 (dashed lines). The colours red, blue, magenta and orange indicate 
	the F, FO, SO, TO pulsation modes,respectively, of the previous relations. 
	Green bow-tie symbols shows the 23 SX~Phe found in Sculptor dSph in this work. 
	They were plotted assuming a true distance modulus of 19.62 mag 
	(\citetalias{MartinezVazquez2015}) and a reddening of 0.018 mag 
	\citep{Pietrzynski2008} plus the reddening law from \citet{Cardelli1989}.}
	\label{fig:pl_sxp}
\end{figure}

Figure~\ref{fig:pl_sxp} shows the absolute \bb-band and \vv-band PL
relations. The latter is used as a consistency check of the pulsation mode
chosen. For the metallicity Z=0.001(Z=0.0001)---solid(dashed) lines---we assign the
following modes: 0(0) F, 16(17) FO, 6(5) SO, 0(0) TO. Both metallicities provide
similar pulsation modes for the population of SX~Phe stars. Furthermore, we note 
that the spread around the PL relations with such a classification is very small, 
i.e., $\sigma_{FO}$=0.02 mag and $\sigma_{SO}$=0.05 mag. However, we note 
that mode classification in this type of variable star is still uncertain. 
For example, the empirical relations of \citet{Pych2001}, with slope and zero-point 
similar to those from the theoretical models of \citet{Fiorentino2015b}, provide 
the same modes. On the other hand, the empirical relations such as, e.g.,
\citet{McNamara2011} (their eq. 5a and 5b) instead suggest that 15(13) are F 
and 5(5) are FO; this is due to the relations having similar slopes but a zero-point 
offset of $\sim$0.3. Further studies of SX~Phe stars targeting nearby galaxies 
(as, e.g., \citealt{Coppola2015} in Carina) are warranted in order to help to 
clarify the situation.

\begin{figure}
	\hspace{-1.7cm}
	\includegraphics[scale=0.6]{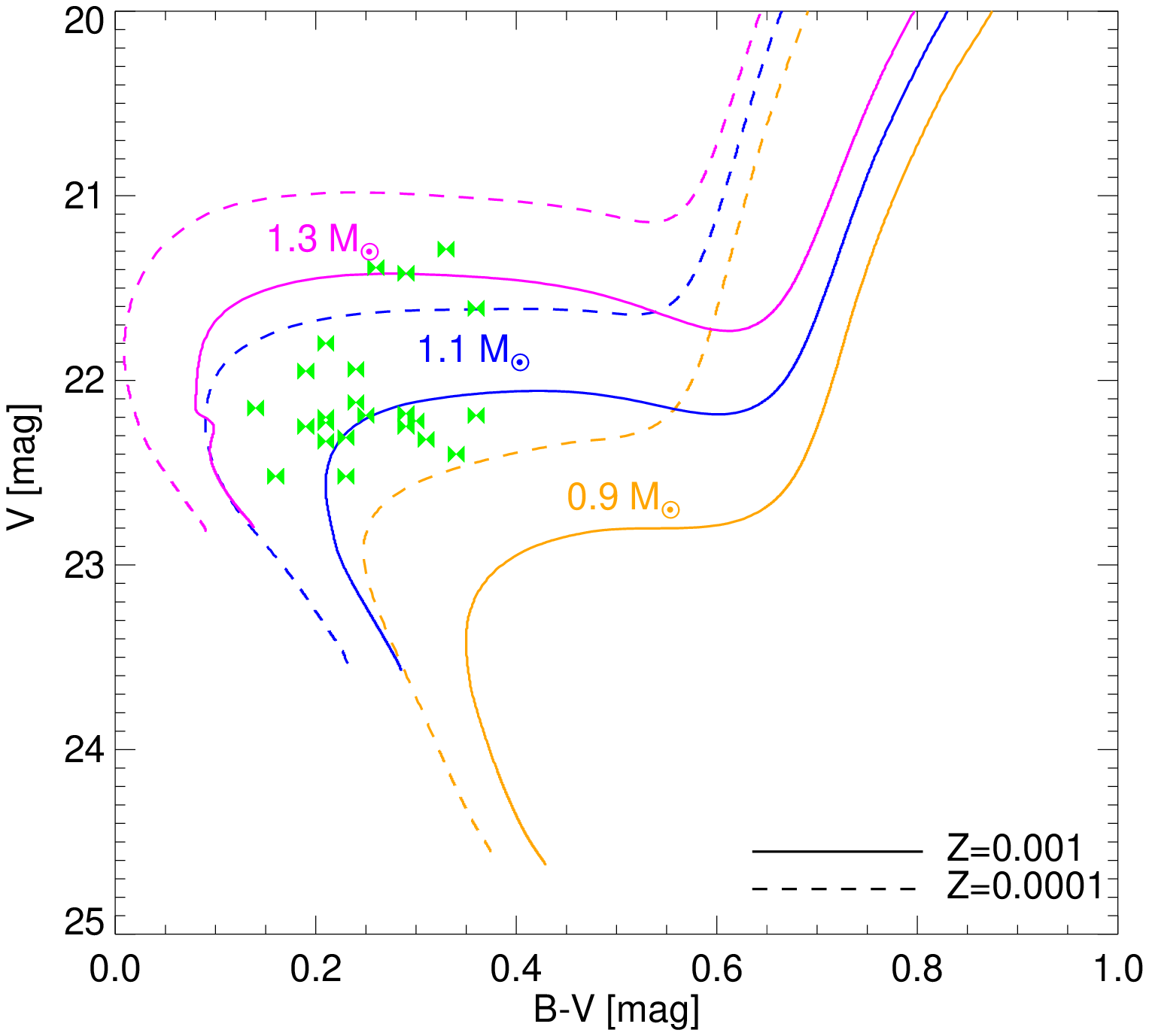}
	\caption{Distribution of the 23 SX~Phe stars in Sculptor dSph in the optical 
	(\bmv, \vv) CMD (with the same colour code as in Fig.~\ref{fig:cmd}). 
	We have represented the standard scaled-solar evolutionary 
	tracks from the BaSTI library \citep{Pietrinferni2004} for different metallicities 
	Z=0.001 (solid lines) and Z=0.0001 (dashed lines), and for the masses 0.9 M$_{\sun}$ 
	(orange lines), 1.1 M$_{\sun}$ (blue lines) and 1.3 M$_{\sun}$ (magenta lines).}
	\label{fig:sxp_tracks}
\end{figure}

Once the pulsation mode is adopted for each SX~Phe star, we can use the
Mass-Period-Luminosity-Metallicity relations (see Table~6 in
\citealt{Fiorentino2015b}) for each selected pulsation mode to estimate the
masses using \bb-band mass dependent PL relations. 
We will adopt the pulsation modes derived from the theoretical relations for
consistency. The mean pulsation masses for
Z=0.001 are: $<$M$_{FO}>$=1.05 M$_{\sun}$ ($\sigma$=0.01), $<$M$_{SO}>$=1.17
M$_{\sun}$ ($\sigma$=0.02). For Z=0.0001, they are: $<$M$_{FO}>$=0.91 M$_{\sun}$
($\sigma$=0.01), $<$M$_{SO}>$=1.01 M$_{\sun}$ ($\sigma$=0.01). These predicted
masses are in good agreement with the evolutionary stellar models (BaSTI,
\citealt{Pietrinferni2004}) as shown in Figure~\ref{fig:sxp_tracks}. 
Incidentally, we note that the age of a single star of such a mass is about 4 Gyr.
Interestingly, in the star-formation history of {\itshape fast} galaxies it is 
common to detect a peak of star formation at this age, which is interpreted
as the contribution of the BSS \citep{Monelli2010a,Monelli2010b,Monelli2016}.

\section{Other variable stars}\label{sec:others}

	\subsection{Peculiar HB variable stars}\label{sec:peculiar}

\begin{figure}
	\hspace{-1cm}
	\includegraphics[scale=0.5]{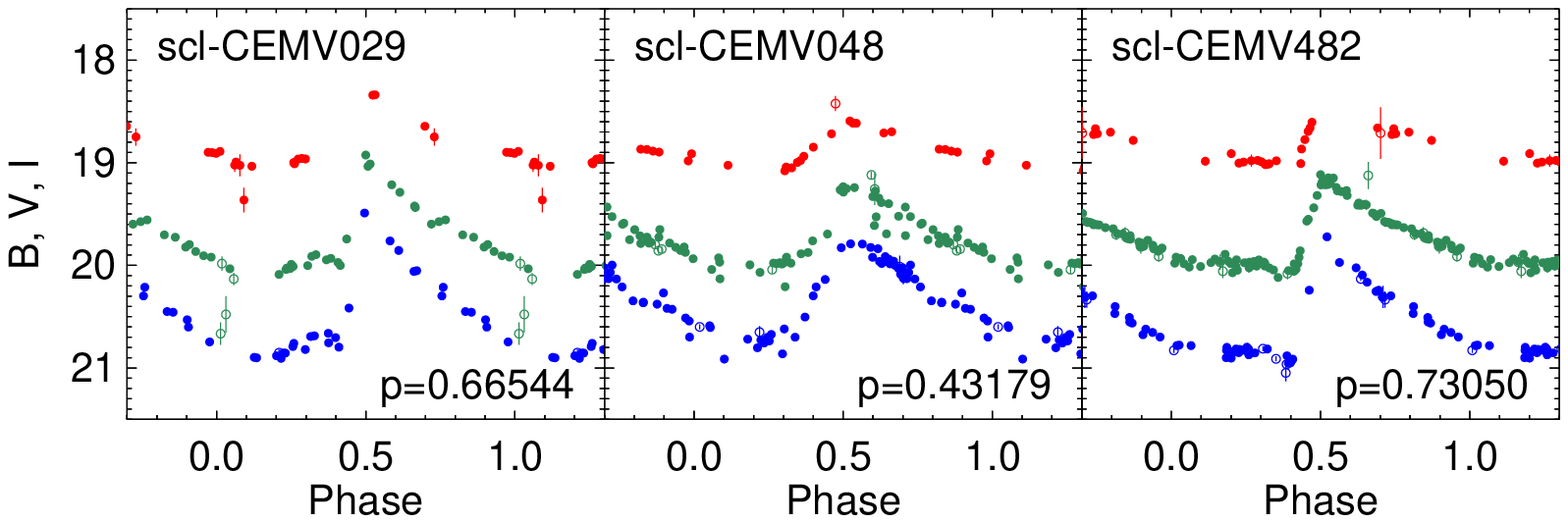}
	\vspace{-9cm}
	\caption{Sample of light curves of the peculiar variable stars in the \bb~
		(blue), \vv~ (green) and \ii~ (red) bands, phased with the period in days
		given  in the lower right-hand corner of each panel. The name of the
		variable is  given in the left-hand corner of each panel. Open symbols
		show the bad data points, i.e, with errors larger than 3$\sigma$ above
		the mean error of a given star; these were not used in the calculation
		of the period and mean magnitudes. For clarity, the \bb~ and \vv~ light curves
		have been shifted by 0.4 mag down- and upward, respectively.}
	\label{fig:peculiar_lcv}
\end{figure}

We have identified three peculiar variable stars, with periods and LCs similar 
to those of RRL, but located quite above the HB. They are $\sim0.3$ mag brighter 
than the brightest RRL stars of the \textit{full} sample, but $\sim0.6$ mag fainter 
than the faintest ACs. The properties of these stars, which we could not classify
convincingly -- see below -- are summarized in Table~\ref{tab:peculiar}, while their 
LCs are presented in Fig.~\ref{fig:peculiar_lcv}.

From an inspection of the CMD these stars are unlikely to be ACs, which are typically 
more than 1 mag brighter that the HB in the $V$ band. Their location suggests they are 
either RRL affected by blending or stars evolving off from the blue part of the HB, 
i.e., stars on the verge to become BL~Herculis (BL~Her) variables. The first hypothesis 
is not convincing since these objects are located far from the galaxy centre 
(see Fig.~\ref{fig:spatial}). Furthermore their amplitudes (see Table~\ref{tab:peculiar}) 
and their amplitude ratios (A$_B$/A$_V \sim 1.34$ and A$_V$/A$_I \sim 1.57$) are similar 
to that expected for normal RRL stars \citep{DiCriscienzo2011} thus not supporting 
a possible blend effect. The second hypothesis is also unlikely, since BL~Her typically 
have periods >1 day \citep[e.g.,][]{Wallerstein2002,Soszynski2010,Marconi2011,Soszynski2015}, 
while these have periods < 0.8 day. Besides, according to the current evolutionary scenario 
the time spent within the Instability Strip by a star with such a luminosity after the 
central Helium burning is very short ($<<$10 Myr; e.g., \citealt{Pietrinferni2004}), 
which means that only in systems with a prominent extreme BHB can we expect to observe 
BL~Her stars (e.g. \citealt{Maas2007}). 

We therefore cannot classify these variables convincingly, but note that similar peculiar 
objects have been detected in other galaxies (e.g., Carina: \citealt{DallOra2003,Coppola2013}, 
Cetus and Tucana: \citealt{Bernard2009}).

\begin{table*}
\begin{scriptsize}
\centering
\caption{Parameters of the three \textit{peculiar} variable stars in Sculptor dSph.} 
\label{tab:peculiar}
  \begin{tabular}{ccccccccccccccc} 
 \hline
CEMV+2016 & Original & Alternative & RA & DEC & Period & <\bb> & <\vv> & <\ii> & A$_{\bb}$ & A$_{\vv}$ & A$_{\ii}$ & Q1 & Q2 \\
name & name & name & (J2000) & (J2000) & (current) & & & & & & & & & \\ 
\hline 
scl-CEMV029   &      ---	&	  ---	 &  00 58 31.92  &  -33 58 19.6  &    0.665435    &  19.949  & 19.663  & 19.135  &  1.433  &  1.114  &  0.648  &  0  &  1   \\
scl-CEMV048   &      V372	&	  ---	 &  00 58 50.94  &  -33 49 31.6  &    0.4317936   &  19.885  & 19.699  & 19.269  &  1.021  &  0.724  &  0.493  &  2  &  0	  \\
scl-CEMV482   &      V152	&	  ---  &  01 00 36.18  &  -33 45 40.7  &    0.7305012   &  20.068  & 19.716  & 19.185  &  1.245  &  0.931  &  0.615  &  0  &  1   \\
\hline
\end{tabular}
\begin{tablenotes}
\item See the caption of Table~\ref{tab:rrl} for a description of the Q1 and Q2 parameters.
\end{tablenotes}
\end{scriptsize}
\end{table*}

%

	\subsection{Eclipsing binaries}\label{sec:eclipsing}

\begin{figure}
	\hspace{-1cm}
	\includegraphics[scale=0.5]{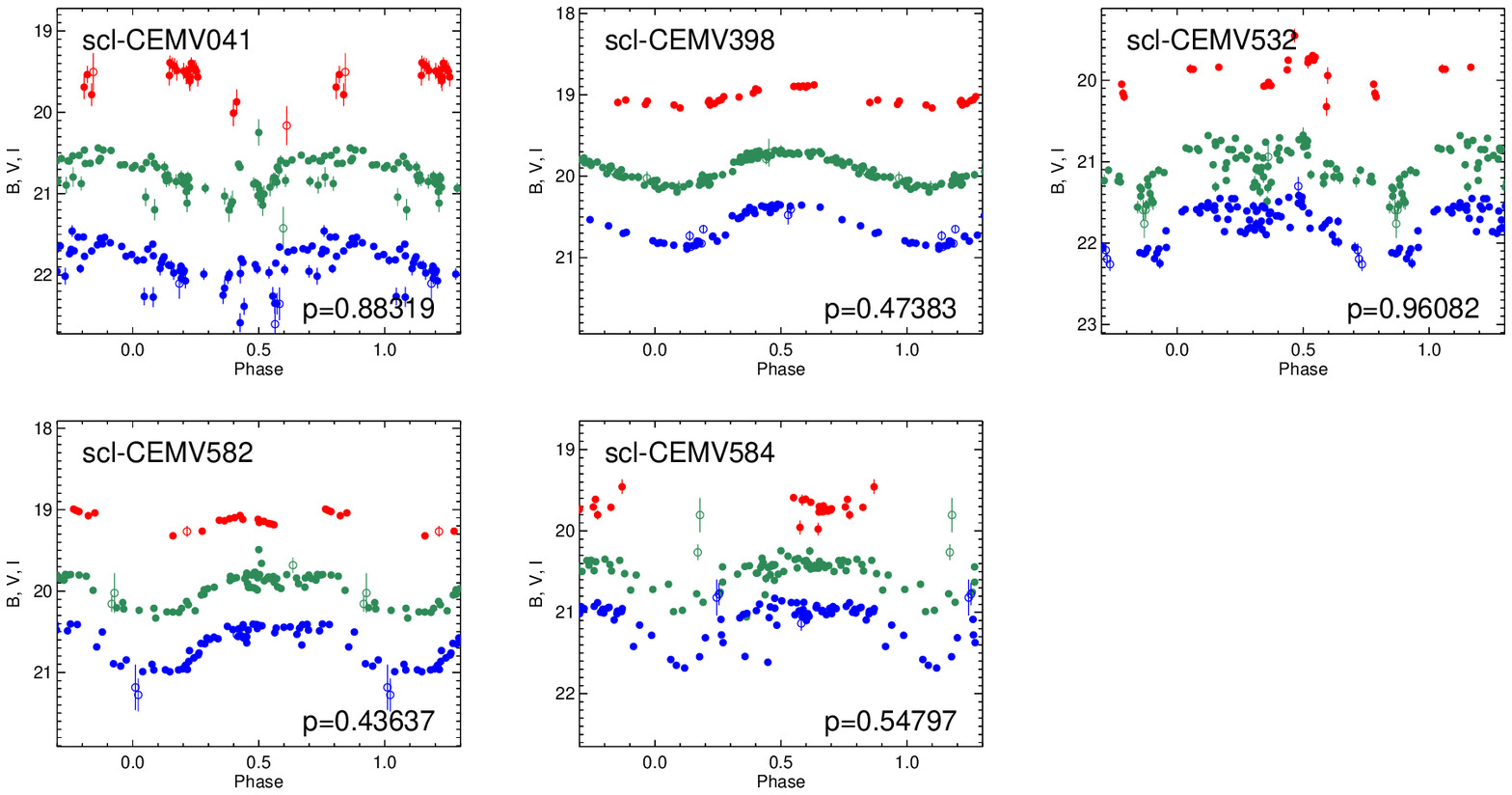}
	\vspace{-7cm}
	\caption{Sample of light curves of the eclipsing binary stars in the \bb~
		(blue), \vv~ (green) and \ii~ (red) bands, phased with the period in days given in the
		lower right-hand corner of each panel. The name of the variable is given in the
		left-hand corner of each panel. Open symbols show the bad data points, i.e, with
		errors larger than 3$\sigma$ above the mean error of a given star; these were
		not used in the calculation of the period and mean magnitudes. For clarity, the
		\bb~ and \ii~ light curves have been shifted by 0.4 mag down- and upward,
		respectively.}
	\label{fig:ebin_lcv}
\end{figure}

Five eclipsing binaries have been detected. The LCs for five of them are shown in
Fig.~\ref{fig:ebin_lcv}, and the properties are summarized in Table~\ref{tab:ebin}.
It is worth noting that, for scl-CEMV398, despite the fact that its LC is very similar to one RRc, we 
classified this star as eclipsing binary based on both the flattening of the brighter part 
of its LC and on the period (P=0.474 d, unusually long for a RRc-type star).
Note also the classification provided by \citet{Clementini2005}, who classified it as ``suspected binary system''.\\

\begin{table*}
\begin{scriptsize}
\centering
\caption{Parameters of the eclipsing binary stars in Sculptor dSph.} 
\label{tab:ebin}
  \begin{tabular}{ccccccccccccccc} 
 \hline
CEMV+2016 & Original & Alternative & RA & DEC & Period & <\bb> & <\vv> & <\ii> & A$_{\bb}$ & A$_{\vv}$ & A$_{\ii}$ & Q1 & Q2 \\
name & name & name & (J2000) & (J2000) & (current) & & & & & & & & & \\ 
\hline 
scl-CEMV041   &      V247	  &	  ---	  &  00 58 47.11  &  -33 50 45.2  &    0.883193:   &  21.383  & 20.688  & 20.051  &  0.332  &  0.315  &  0.600  &  1  &  0  \\ 
scl-CEMV398   &      K3710	&	  ---	  &  01 00 18.65  &  -33 45 35.1  &    0.4738299   &  20.190  & 19.896  & 19.418  &  0.500  &  0.398  &  0.263  &  0  &  1  \\ 
scl-CEMV532   &      V410	  &	  ---	  &  01 00 51.82  &  -33 47 55.4  &    0.960817:   &  21.266  & 20.901  & 20.236  &  0.414  &  0.526  &  0.645  &  1  &  0  \\ 
scl-CEMV582   &      V315	  &	  ---	  &  01 01 18.33  &  -33 43 42.8  &    0.4363745:  &  20.214  & 19.978  & 19.518  &  0.504  &  0.335  &  0.063  &  1  &  0  \\ 
scl-CEMV584   &      ---	  &	  ---	  &  01 01 19.77  &  -33 36 21.0  &    0.547969:   &  20.728  & 20.539  & 19.633  &  0.383  &  0.294  &  0.974  &  1  &  0  \\ 
\hline
\end{tabular}
\begin{tablenotes}
\item See the caption of Table~\ref{tab:rrl} for a description of the Q1 and Q2 parameters.
\end{tablenotes}
\end{scriptsize}
\end{table*}


	\subsection{Field variable stars}\label{sec:field}

\begin{figure}
	\hspace{-1cm}
	\includegraphics[scale=0.5]{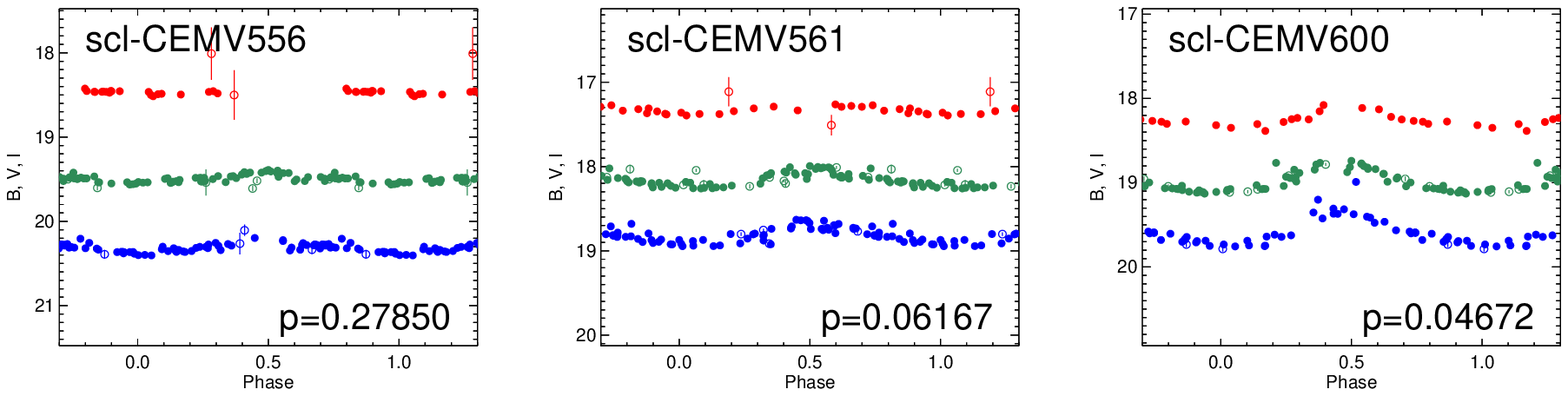}
	\vspace{-10cm}
	\caption{Sample of light curves of the field variable stars in the \bb~
		(blue), \vv~ (green) and \ii~ (red) bands, phased with the period in days given in the
		lower right-hand corner of each panel. The name of the variable is given in the
		left-hand corner of each panel. Open symbols show the bad data points, i.e, with
		errors larger than 3$\sigma$ above the mean error of a given star; these were
		not used in the calculation of the period and mean magnitudes. For clarity, the
		\bb~ and \ii~ light curves have been shifted by 0.4 mag down- and upward,
		respectively.}
	\label{fig:field_lcv}
\end{figure}

We have identified three variable stars compatible with being foreground field
stars. Their LCs are shown in Fig.~\ref{fig:field_lcv}, and the properties are
summarized in Table~\ref{tab:field}. Two are compatible with being field
$\delta$Scuti stars.

\begin{table*}
\begin{scriptsize}
\centering
\caption{Parameters of the field variable stars found in Sculptor dSph.} 
\label{tab:field}
  \begin{tabular}{ccccccccccccccc} 
 \hline
CEMV+2016 & Original & Alternative & RA & DEC & Period & <\bb> & <\vv> & <\ii> & A$_{\bb}$ & A$_{\vv}$ & A$_{\ii}$ & Q1 & Q2 \\
name & name & name & (J2000) & (J2000) & (current) & & & & & & & & & \\ 
\hline 
scl-CEMV556   &      ---	&	  ---	 &  01 01 05.12  &  -33 46 00.9  &    0.2784994   &  19.905  & 19.488  & 18.866  &  0.129  &  0.108  &  0.031  &  4  &  0  \\
scl-CEMV561   &      ---	&	  ---	 &  01 01 06.68  &  -33 54 46.9  &    0.06166818  &  18.408  & 18.150  & 17.723  &  0.234  &  0.197  &  0.092  &  4  &  0  \\
scl-CEMV600   &      ---	&	  ---	 &  01 01 31.04  &  -33 25 12.4  &    0.04672141  &  19.151  & 18.963  & 18.651  &  0.409  &  0.316  &  0.188  &  3  &  0  \\
\hline
\end{tabular}
\begin{tablenotes}
\item See the caption of Table~\ref{tab:rrl} for a description of the Q1 and Q2 parameters.
\end{tablenotes}
\end{scriptsize}
\end{table*}

	\subsection{Likely candidates}\label{sec:cpc}

A sizeable sample of 37 LPV and 23 probable variable stars has been detected 
in Sculptor. For the case of the LPV stars, due to their long periods, a large
number of phase points together with an appropriately long temporal coverage are needed
to define the shape of their LCs. Despite our large data set, we have not been able to 
characterize the properties of these stars. We consider a star to be a candidate 
LPV when we obtain concordant magnitude measurements within individual nights and 
individual observing runs, but average results from observing runs separated by many 
months or many years are highly discrepant. 

For the case of the probable variables, the
insufficient data and the inability to achieve a good light curve makes the classification
and the period determination of such stars difficult to do. Based on these facts, we named 
them ``likely candidates''. Table~\ref{tab:cpc} shows the list of these stars. We note that:
{\it i)} 31 out of 37 LPV stars are located near the tip of the RGB; and 
{\it ii)} out of the 23 probable variable stars, 4 could be possibly eclipsing binaries, and 2 
could be RRc stars.

\begin{table*}
\begin{scriptsize}
\centering
\caption{Parameters of the candidates variable stars with problems of clasification in Sculptor dSph.} 
\label{tab:cpc}
  \begin{tabular}{cccccl} 
\hline
CEMV+2016 & Original & Alternative & RA & DEC & Note \\
name & name & name & (J2000) & (J2000) &  \\ 
\hline
scl-CEMV003   &      V579	&	  ---  &  00 56 29.94  &  -33 56 03.9  &   probable variable							  \\
scl-CEMV004   &      V503	&	  ---  &  00 56 32.41  &  -33 54 52.4  &   probable variable								  \\
scl-CEMV007   &      V443	&	  ---  &  00 56 57.55  &  -33 59 20.1  &   probable variable								  \\
scl-CEMV009   &      ---	&	  ---  &  00 57 14.08  &  -33 48 19.6  &   probable variable; probable RRc but unable to determine period				  \\
scl-CEMV011   &      V374	&	  ---  &  00 57 37.90  &  -33 58 57.0  & 	 probable variable									  \\
scl-CEMV013   &      ---	&	  ---  &  00 57 47.20  &  -33 31 36.7  &   probable variable									  \\
scl-CEMV017   &      V546	&	  ---  &  00 57 59.79  &  -33 55 58.1  &   probable variable								  \\
scl-CEMV023   &      V581	&	  ---  &  00 58 15.63  &  -33 47 57.4  &   probable variable; possible eclipsing binary							  \\
scl-CEMV075   &      ---	&	  ---  &  00 59 04.80  &  -33 38 44.2  &   LPV								  \\
scl-CEMV082   &      ---	&	  ---  &  00 59 08.58  &  -33 41 52.7  &   LPV							  \\
scl-CEMV101   &      ---	&	  ---  &  00 59 15.76  &  -33 42 48.6  &   LPV							  \\
scl-CEMV105   &      ---	&	  ---  &  00 59 16.92  &  -33 40 10.6  &   LPV							  \\
scl-CEMV117   &      V356	&	  ---  &  00 59 20.53  &  -33 14 44.2  &   probable variable								  \\
scl-CEMV121   &      ---	&	  ---  &  00 59 23.33  &  -33 23 48.3  &   probable variable; possible eclipsing binary							  \\
scl-CEMV136   &      ---	&	  ---  &  00 59 27.67  &  -33 40 35.6  &   LPV							  \\
scl-CEMV140   &      ---	&	  ---  &  00 59 28.28  &  -33 42 07.4  &   LPV							  \\
scl-CEMV159   &      ---	&	  ---  &  00 59 33.95  &  -33 38 37.3  &   LPV							  \\
scl-CEMV164   &      K274	&	  ---  &  00 59 35.36  &  -33 44 09.4  &   LPV							  \\
scl-CEMV168   &      ---	&	  ---  &  00 59 36.13  &  -33 44 33.4  &   LPV							  \\
scl-CEMV181   &      V115	&  K737  &  00 59 37.98  &  -33 42 54.1  &   probable variable			  \\
scl-CEMV194   &      V204	&	  ---  &  00 59 41.48  &  -33 51 36.8  &   probable variable; multimode?		  \\
scl-CEMV196   &      V539	&	  ---  &  00 59 42.42  &  -33 46 57.9  &   probable variable; possible eclipsing binary, but unable to determine period				  \\
scl-CEMV203   &      ---	&	  ---  &  00 59 43.15  &  -33 56 46.7  &   LPV							  \\
scl-CEMV221   &      K687	&	  ---  &  00 59 46.40  &  -33 41 23.4  &   LPV							  \\
scl-CEMV228   &      ---	&	  ---  &  00 59 49.18  &  -33 58 25.2  &   LPV							  \\
scl-CEMV241   &      ---	&	  ---  &  00 59 52.27  &  -33 44 54.7  &   LPV								  \\
scl-CEMV244   &      ---	&	  ---  &  00 59 53.00  &  -33 39 19.0  &   LPV								  \\
scl-CEMV254   &      V97	&	  ---  &  00 59 54.63  &  -33 43 42.6  &   LPV								  \\
scl-CEMV289   &      V544	&	  ---  &  00 59 58.92  &  -33 28 35.0  &   LPV								  \\
scl-CEMV296   &      ---  &	  ---  &  01 00 00.06  &  -33 38 34.3  &   LPV								  \\
scl-CEMV303   &      V551	&	  ---  &  01 00 01.13  &  -33 59 21.3  &   probable variable 			  \\
scl-CEMV343   &      V80	&	  ---  &  01 00 06.15  &  -33 50 36.5  &   LPV							  \\
scl-CEMV346   &      ---	&	  ---  &  01 00 07.56  &  -33 37 03.8  &   LPV								  \\
scl-CEMV368   &      ---	&	  ---  &  01 00 12.12  &  -33 37 25.9  &   LPV							  \\
scl-CEMV372   &    K4780  &	  ---	 &  01 00 13.97  &  -33 37 32.9  &	 probable variable; possible detached eclipsing binary \\ 
scl-CEMV395   &      ---	&	  ---  &  01 00 18.19  &  -33 31 40.4  &   LPV								  \\
scl-CEMV412   &      ---	&	  ---  &  01 00 20.72  &  -33 45 18.0  &   LPV							  \\
scl-CEMV450   &      ---	&	  ---  &  01 00 30.74  &  -33 37 30.5  &   LPV								  \\
scl-CEMV460   &      ---	&	  ---  &  01 00 32.35  &  -33 31 57.8  &   LPV								  \\
scl-CEMV462   &      V575	&	  ---  &  01 00 32.83  &  -33 38 32.4  &   probable variable, period near 1d						  \\
scl-CEMV469   &      ---	&	  ---  &  01 00 34.03  &  -33 39 04.6  &   LPV								  \\
scl-CEMV503   &      ---	&	  ---  &  01 00 41.77  &  -33 51 40.8  &   possible  variable, but unable to determine period; possibly near 1d		  \\
scl-CEMV504   &      ---	&	  ---  &  01 00 42.55  &  -33 35 47.4  &   LPV							  \\
scl-CEMV530   &      ---	&	  ---  &  01 00 50.86  &  -33 45 05.3  &   LPV								  \\
scl-CEMV534   &      ---	&	  ---  &  01 00 52.13  &  -33 41 27.1  &   probable variable									  \\
scl-CEMV550   &      ---	&	  ---  &  01 01 02.87  &  -33 38 52.2  &   LPV								  \\
scl-CEMV566   &      ---	&	  ---  &  01 01 08.54  &  -33 45 34.9  &   LPV								  \\
scl-CEMV573   &      ---	&	  ---  &  01 01 14.54  &  -33 30 09.2  &   LPV							  \\
scl-CEMV578   &      ---	&	  ---  &  01 01 17.33  &  -33 37 05.1  &   LPV								  \\
scl-CEMV586   &      ---	&	  ---  &  01 01 20.82  &  -33 53 04.7  &   LPV								  \\
scl-CEMV591   &      ---	&	  ---  &  01 01 24.58  &  -33 38 34.5  &   LPV								  \\
scl-CEMV595   &      ---	&	  ---  &  01 01 26.93  &  -33 41 44.7  &   probable variable; probable RRc but unable to determine period				  \\
scl-CEMV607   &      ---	&	  ---  &  01 01 39.09  &  -33 56 38.5  &   LPV							  \\
scl-CEMV608   &      ---	&	  ---  &  01 01 39.58  &  -33 45 04.3  &   LPV								  \\
scl-CEMV613   &      ---	&	  ---  &  01 01 49.41  &  -33 54 10.2  &   LPV								  \\
scl-CEMV615   &      V480	&	  ---  &  01 01 54.54  &  -34 05 21.7  &   probable variable							  \\
scl-CEMV628   &      V530	&	  ---  &  01 02 44.13  &  -33 25 54.4  &   probable variable								  \\
scl-CEMV629   &      V577	&	  ---  &  01 02 45.21  &  -33 46 36.4  &   probable variable							  \\
scl-CEMV631   &      ---	&	  ---  &  01 02 52.67  &  -33 37 26.9  &   probable variable									  \\
scl-CEMV633   &      V587	&	  ---  &  01 03 43.20  &  -34 23 34.3  &   probable variable						  \\
\hline
\end{tabular}
\end{scriptsize}
\end{table*}

	\subsection{Non-variable stars}\label{sec:nonvariable}

The comparison with previous work on variable stars in Sculptor discloses that
we do not detect any trace of variability in a number of sources previously 
catalogued as variable, mostly from the \citet{vanAgt1978} paper, and a few
from the \citet{Kaluzny1995}. The list is presented in Table~\ref{tab:novar}.

\begin{table*}
\begin{scriptsize}
\centering
\caption{Parameters of non-variable stars in Sculptor dSph.} 
\label{tab:novar}
  \begin{tabular}{lcccl} 
\hline 
Original & Alternative & RA & DEC & Note\\
name & name & (J2000) & (J2000) &  \\ 
\hline 
V573		&   ---	  &  00 56 38.63  &  -33 25 23.7   &   not variable?							      \\			 
V397		&   ---	  &  00 57 24.12  &  -33 38 02.5   &   not variable?							      \\  
V580		&   ---	  &  00 57 33.20  &  -33 52 00.2   &   not variable?							      \\  
V596		&   ---   &  00 57 49.52  &  -33 32 23.3   &   not variable?							      \\  
V559		&   ---   &  00 58 12.31  &  -33 37 50.4   &   not variable?							      \\  
V547		&   ---	  &  00 58 36.48  &  -34 02 11.9   &   not variable?							      \\  
V252		&   ---	  &  00 58 45.65  &  -33 32 12.8   &   not variable							      \\  
V251		&   ---	  &  00 59 00.13  &  -33 38 50.9   &   not variable							      \\  
V311		&   ---	  &  00 59 21.44  &  -33 48 48.6   &   not variable?							      \\  
V332		&   ---	  &  00 59 32.85  &  -33 57 44.4   &   not variable							      \\  
V554		&   ---   &  00 59 35.22  &  -33 29 04.3   &   not variable?							      \\  
K403		&   ---	  &  00 59 36.61  &  -33 46 05.7   &   not variable							      \\  
V416		&   ---	  &  00 59 44.35  &  -33 55 19.3   &   not variable?							      \\  
V200		&   ---	  &  00 59 47.20  &  -33 33 37.0   &   not variable?							      \\  
V382		&   ---	  &  00 59 57.66  &  -33 46 56.5   &   not variable?							      \\  
V370		&   ---	  &  01 00 06.99  &  -33 52 25.7   &   not variable?							      \\  
V173		&   ---	  &  01 00 17.17  &  -33 56 06.4   &   not variable							      \\  
V60		  &   ---	  &  01 00 21.78  &  -33 39 07.7   &   no bright star here; estimated position is 11.4 arcseconds from K4313   \\  
V245		&   ---	  &  01 00 24.43  &  -33 25 09.7   &   not variable							      \\  
V558		&   ---	  &  01 00 24.75  &  -33 51 06.6   &   not variable?							      \\  
V557		&   ---   &  01 00 26.14  &  -33 41 07.3   &   not variable?							      \\  
V54		  &   ---	  &  01 00 28.99  &  -33 51 52.5   &   not variable							      \\  
K5081		&   ---	  &  01 00 30.78  &  -33 46 27.1   &   not variable?							      \\  
V45		  &   ---	  &  01 00 39.78  &  -33 52 21.9   &   not variable							      \\  
V532		&   ---	  &  01 00 43.79  &  -33 29 08.1   &   not variable?							      \\  
V424		&   ---	  &  01 00 47.84  &  -33 58 54.2   &   not variable?							      \\  
V340		&   ---   &  01 01 21.98  &  -33 42 22.6   &   not variable?							      \\  
V571		&   ---	  &  01 02 36.69  &  -33 30 07.1   &   not variable?							      \\  
V307		&   ---	  &  01 03 23.63  &  -33 44 16.8   &   not variable?							      \\  
\hline
\end{tabular}
\end{scriptsize}
\end{table*}

\section{Discussion: The fast early chemical evolution of Sculptor and Tucana}
\label{sec:discussion}

In this work, we presented the most complete and updated catalogue of variable
stars in the Sculptor dSph. This is so far one of the largest RRL star samples
in external galaxies of similar morphological type. As demonstrated in 
\citetalias{MartinezVazquez2015}, the combination of a large sample together
with high quality and multi-band photometry, allowed us to set tight constraints
on the metallicity distribution of the old stellar component, and revealed the
presence of a significant metallicity spread within the RRL star population (t
$>$ 10 Gyr). This implies that Sculptor underwent substantial chemical enrichment
fast enough to be imprinted in the population we observe today as RRL stars.
This manifests itself through a large luminosity spread of the RRL stars 
($\sim$0.35 mag) that is inconsistent with the evolution of a monometallic
population \citepalias{MartinezVazquez2015}. Moreover, we showed that, when
splitting the sample of RRL stars according to their luminosity relative to
the mean $<\vv>$ mag, the brighter and the fainter subsamples follow
different spatial distributions, the latter being more centrally concentrated
than the former. If we interpret the bright and faint components in terms of
metallicity, the latter is consistent with being more metal-rich. This result
is consistent with he spectroscopic results of \citet{Tolstoy2004}. This
implies that, in agreement with what is generally found in dwarf galaxies, 
the youngest and chemically more evolved population is located in the 
innermost regions, surrounded by a more uniform older and more metal-poor one, 
suggesting a metallicity gradient. In the
present work (see \S~\ref{sec:bailey}) we showed that the Bailey diagram is
quite complex, with stars populating the region around and intermediate to the
typical Oo-I and Oo-II loci. Interestingly, we found that the bright,
metal-poor, more extended population preferentially follows the location of the
Oo-II (typically more metal-poor) GCs, while the faint, more metal-rich, more
centrally concentrated RRL stars have a distribution closer to that of Oo-I
clusters.

\begin{figure}
	\includegraphics[scale=0.5]{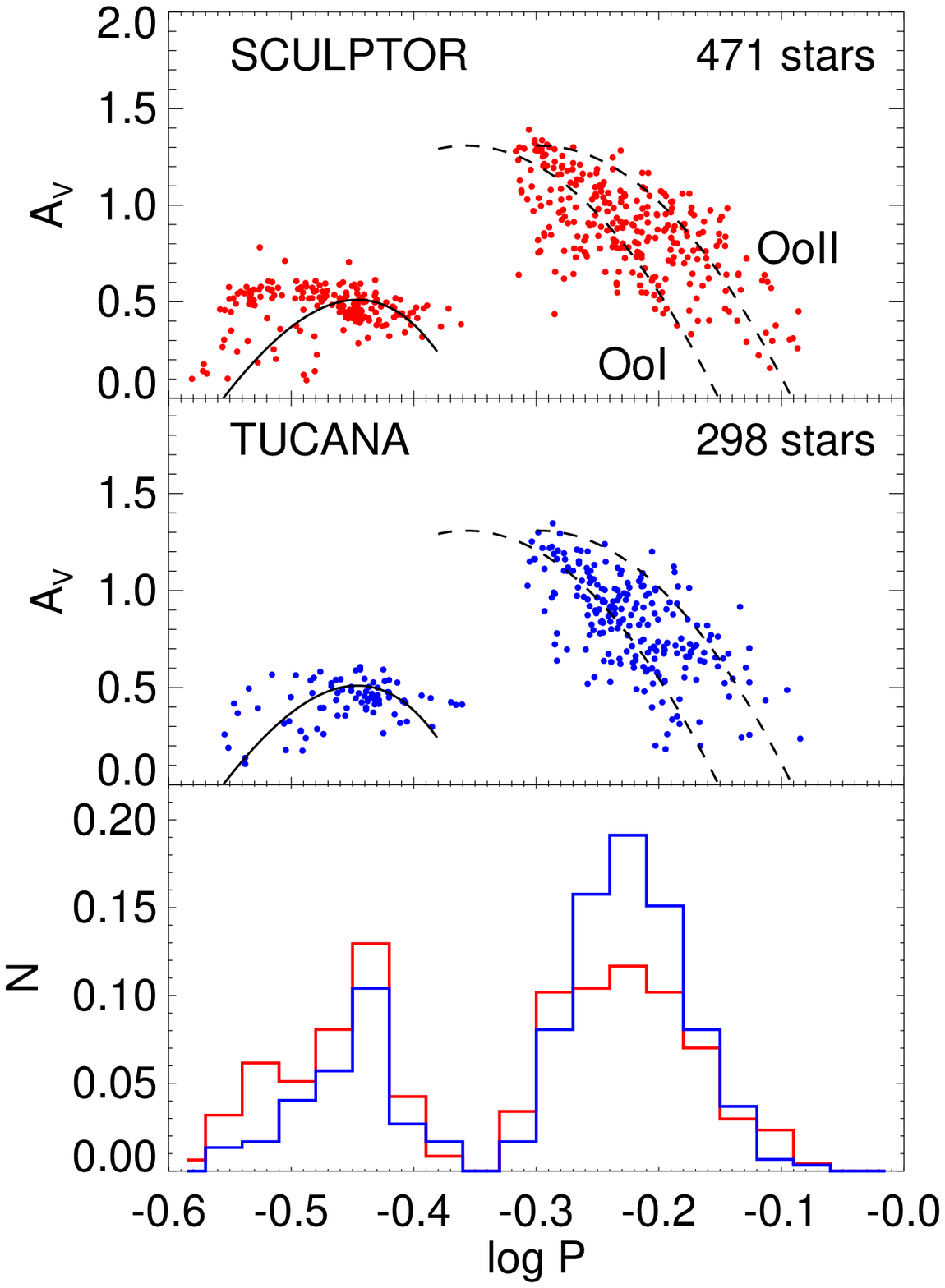}
	\caption{Top. Period-\vv~amplitude diagram for Sculptor (top) and Tucana
		(middle, \citealt{Bernard2009}). For clarity, RRd stars do not appear in
		these diagrams, which only consider stars with well defined pulsational
		properties. The bottom panel shows the normalized distribution of RRab and
		RRc stars in both galaxies (red: Sculptor, blue: Tucana).}
	\label{fig:scl_tuc_bailey}
\end{figure}

A surprisingly similar empirical result was found in the Tucana dSph.
\citet{Bernard2008} disclosed that the pulsational properties of the RRL stars
in this galaxy trace a spatial gradient in the metallicity of its individual 
stars, which therefore must have appeared very early on in the history of this 
galaxy. As in Sculptor, fainter RRLs are more centrally concentrated than the brighter 
RRLs. This was the first time that a spatial variation of pulsational properties was
observed in a dwarf galaxy, thanks to the large spatial coverage and number of
variables discovered. Fig.~\ref{fig:scl_tuc_bailey} compares the Bailey
diagram and the period distribution of the RRL stars populations in Sculptor and
Tucana. The distribution of the RRab-type shows the same general
characteristics, with a large period dispersion at fixed amplitude. This is
reflected in the normalized histogram in the lower panel, which clearly shows
that the period coverage is essentially the same.

Indeed, the old population of the two systems presents striking similarities,
which are unique among the low-mass galaxies of the LG. No other dSph
investigated so far shows such clear evidence of chemical evolution
imprinted in their population of RRL stars. This may suggest that the early
conditions of the two galaxies, and in particular the mass and star
formation histories may have been similar, in order to explain that both systems
were able to retain enough nucleosynthesis products to provide early enrichment as observed 
today. Nonetheless, Sculptor and Tucana also exhibit important differences. 

At present, Tucana is a very isolated dSph at the edge of the Local Group 
($\sim$ 870 kpc from the MW versus $\sim$ 84 kpc for Sculptor). 
Given its current location in the Local Group and its relatively high 
recession velocity, Tucana seems to have been an
isolated Local Group galaxy during the majority of its lifetime, except perhaps
for a close encounter with the MW or M31 at early epochs \citep{Fraternali2009}.
On the contrary, with an apogalactic distance of 122 kpc and orbital period of
2.2 Gyr \citep{Piatek2006}, Sculptor spent most of its existence within the halo
of the MW. Under these conditions, theoretical investigations indicate that tidal
stripping, stirring, and ram-pressure stripping \citep{Blitz2000, Mayer2006},
as well as the local UV radiation from the primary galaxy \citep{Mayer2007} all act to
remove dark matter and/or baryons from the dwarf, implying that satellite galaxies 
such as Scl may have been up to ten times more massive in the past \citep{Kravtsov2004}. 
However, Sculptor is considerably more massive than Tucana at the present time
(M$_{\vv}$= --11.1 versus --9.6 for Tucana). If Sculptor had been ten times as massive as
we observe it today, the mass-metallicity relation would suggest a
larger increase in metallicity at early times. However, this is in contrast with the lack of
HASP RRL stars, which are solid tracers of an old stellar population more
metal-rich than [Fe/H]$\gtrsim$--1.5. Moreover, \citet{Coleman2005} demonstrated
that beyond the tidal radius (from both photometric  and spectroscopic data)
there is no evidence of extra-tidal structure, suggesting the absence of strong
tidal interaction.  This argues against substantial mass loss along Sculptor's 
history and may suggest that, possibly, Sculptor has quietly and
passively evolved during its revolutions around the Milky Way. The similarity of the 
stellar populations of Sculptor and Tucana, including at early times, provides support 
to the scenario about the origin of the dwarf galaxy types introduced by 
\citet{Gallart2015}, in which dwarf galaxy types may be imprinted by the early 
conditions of formation rather than only being the result of a recent or secular morphological 
transformation driven by environmental effects. In this particular case, both Sculptor and 
Tucana may have formed in the relatively high density environment close to the centre of what 
would become the Local Group, and this would be instrumental in becoming {\it fast} galaxies, 
i.e., galaxies whose star formation history is dominated by en early and short 
star formation event, with little star formation afterwards. The subsequent sustained 
interaction of Sculptor with the Milky Way would have had a minor effect in its star formation 
history and mass characteristics, and thus Sculptor and Tucana remain galaxies with similar 
mass and star formation history today despite the substantially different secular evolution.

\section{Summary and Conclusions}\label{sec:conclusions}

We have presented the largest catalogue (so far) of variable stars in the Sculptor dSph
galaxy. This work is based on the homogeneous photometric analysis of 4,404
images in the \bb, \vv, and \ii~ bands collected over 24 years in 21 different
observing runs and using 6 different telescopes and 7 different instruments, employing 
the same methodologies as the ``Homogeneous Photometry'' series. The main results of this work are: \\

\begin{itemize}
	
\item[$\bullet$] Basic properties (period, amplitude, mean magnitude, position) have been made
available for all the stars, together with the light curves and the finding charts 
(see Appendix~\ref{sec:fchart}).	\\
\item[$\bullet$] In total, we have discovered 147 variable stars in the calibrated portion of the Sculptor
dSph, among which 81 are RRLs, 23 SX~Phe, one \textit{peculiar}, one eclipsing binary and 38 ``likely
candidates'' (31 of them are probable LPV stars, one is a possible eclipsing binary, two are probable RRc and 
four more are possible variables of uncertain type).
\item[$\bullet$] Out of the 634 detected variables in the current work, 334 (301 RRLs) have their periods
identified for the first time, and 354 (320 RRLs) have their pulsation parameters also given for the 
first time.
\item[$\bullet$] We have detected 536 RR Lyrae variable stars. Out of these, 289 are RRab type
stars, 197 are RRc, and 50 stars are suspected RRd double-mode pulsators.
We have discussed the distribution of stars in the Bailey diagram, showing that the
metallicity spread among RRL stars discussed in \citetalias{MartinezVazquez2015} reflects not
only in the luminosity spread of the HB, but also in the broad period distribution that
covers both Oo-I and Oo-II loci as well as the intermediate regions and in the amplitude range for 
a given period;\\
\item[$\bullet$] We have confirmed the existence of four ACs over the surveyed area. We have
derived the pulsation mode and the mass of each of them, which is close to 1.5
M$_{\sun}$ for all. \\
\item[$\bullet$] We have discussed 23 newly discovered SX~Phe stars. Using theoretical models 
developed by \citep{Fiorentino2015b}, and the distance derived from RRLs in 
\citetalias{MartinezVazquez2015}, we classified them as 16(17) FO, 6(5) SO, assuming 
a mean metallicity of Z=0.001(Z=0.0001). On the other hand, the empirical relations by 
\citet{McNamara2011} instead suggest that 15(13) are F and 5(5) are FO. The discrepancy 
come from the zero-point offset ($\sim$0.3 mag) between both relations.
The mean mass derived for them is of about 1 M$_{\sun}$.
If we assume that the entire sample of SX~Phe comes from single star evolution, they might indicate a 
residual star formation $\simeq$4 Gyr ago, or coalescence of very low mass stars.\\
\item[$\bullet$] We also discuss the existence of three \textit{peculiar} variable stars, located
in the region of the CMD between the RRL star and the ACs, which have pulsation
properties inconsistent with other classes of variable brighter than the HB, such as
the BL Her stars. Their nature remains unclear. \\
\item[$\bullet$] Five eclipsing binaries and 37 probable long period variables, and a few (3)
field variable stars were also presented.\\

\end{itemize}

We have discussed the striking similarities between the properties of the old
population in the Sculptor and Tucana dSph galaxies, which are imprinted in the
complex populations of their RRL stars.
Despite the large spatial coverage of the present work ($\sim$2.5 deg$^2$), a complete 
investigation of stellar variability over the full tidal radius of Sculptor is still lacking.
Moreover, a direct confirmation of the metallicity spread among the RRL 
population through spectroscopic follow-up could provide further insight into the
early chemical evolution of this galaxy.
The future development of this project does require new precise and radial velocity measurements
of both RG and HB stars to assess whether the two different populations show different kinematics 
and different chemical enrichment histories as recently suggested by (\citealt{Fabrizio2015}, 
2016 AJ accepted) for Carina dSph. Note that the quoted approach took advantage of the $c_{UBI}$ 
index to separate stellar populations along the RGB. This means that deep and accurate U-band 
photometry is also urgently required.

\section*{Acknowledgments}

We thank the referee for his/her comments, careful reading and detailed report.
CEM-V is grateful to the Rome Observatory and the Physics Department of 
the Tor Vergata University where part of this work has been carried out. 
CEM-V acknowledges the support by the ULL through a grant funded by the 
Agencia Canaria de Investigaci\'on, Innovaci\'on y Sociedad de la Informaci\'on
and co-funded by the Fondo Social Europeo (FSE) under the framework of
Programa Operativo de Canarias (POC 2007-2013). 
This research has been supported by the Spanish Ministry of Economy and 
Competitiveness (MINECO) under the grant (project reference AYA2014-56795-P).
EJB acknowledges support from the CNES postdoctoral fellowship program.
GF has been supported by the Futuro in Ricerca 2013 (grant RBFR13J716).




\bibliographystyle{mnras}

 \newcommand{\noop}[1]{}

\appendix
\section{Finding chart}\label{sec:fchart}
\begin{figure*}
	\includegraphics[scale=0.6]{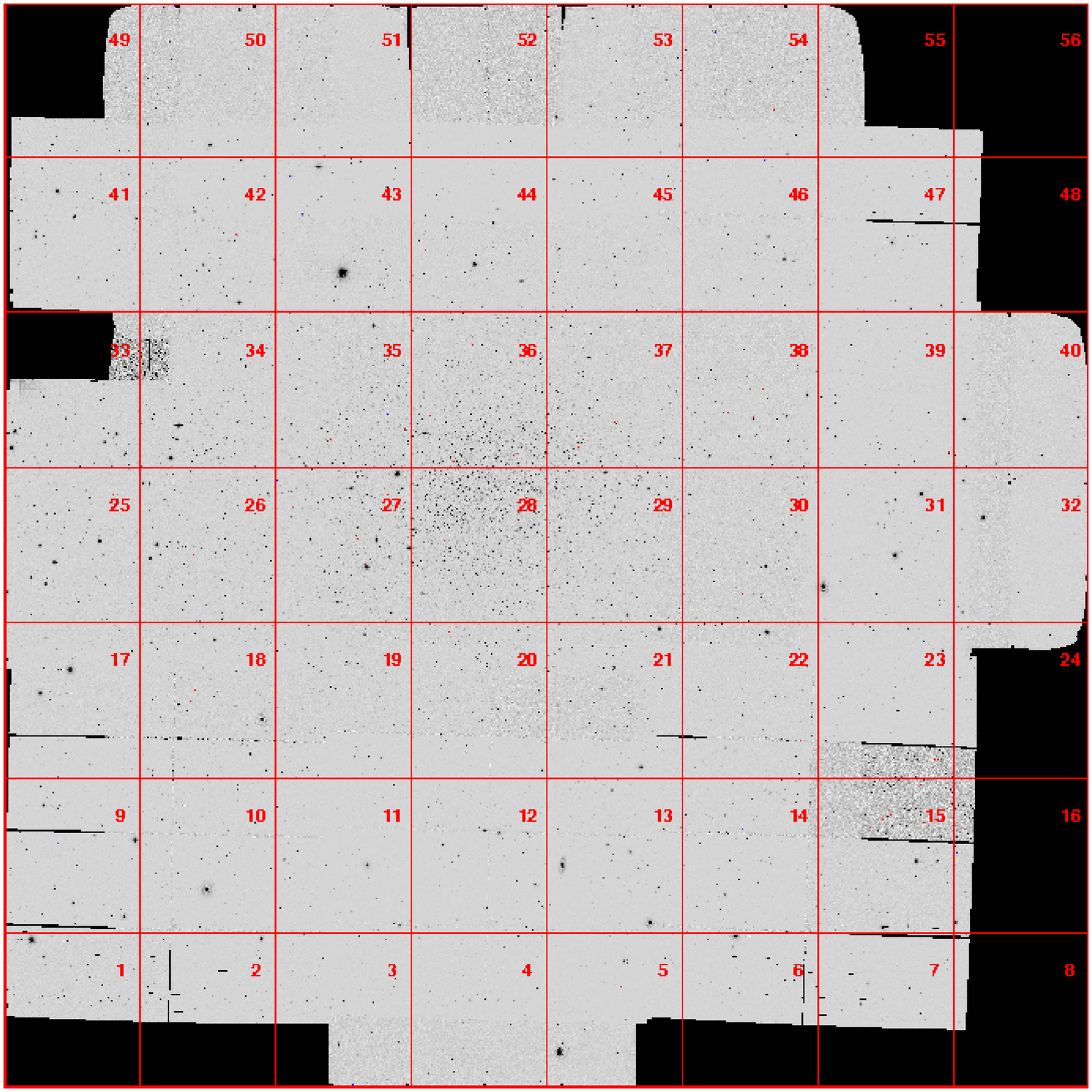}
	\caption{Mosaic of Sculptor.}
\label{fig:mosaic}	
\end{figure*}

\begin{figure*}
	\includegraphics[scale=0.6]{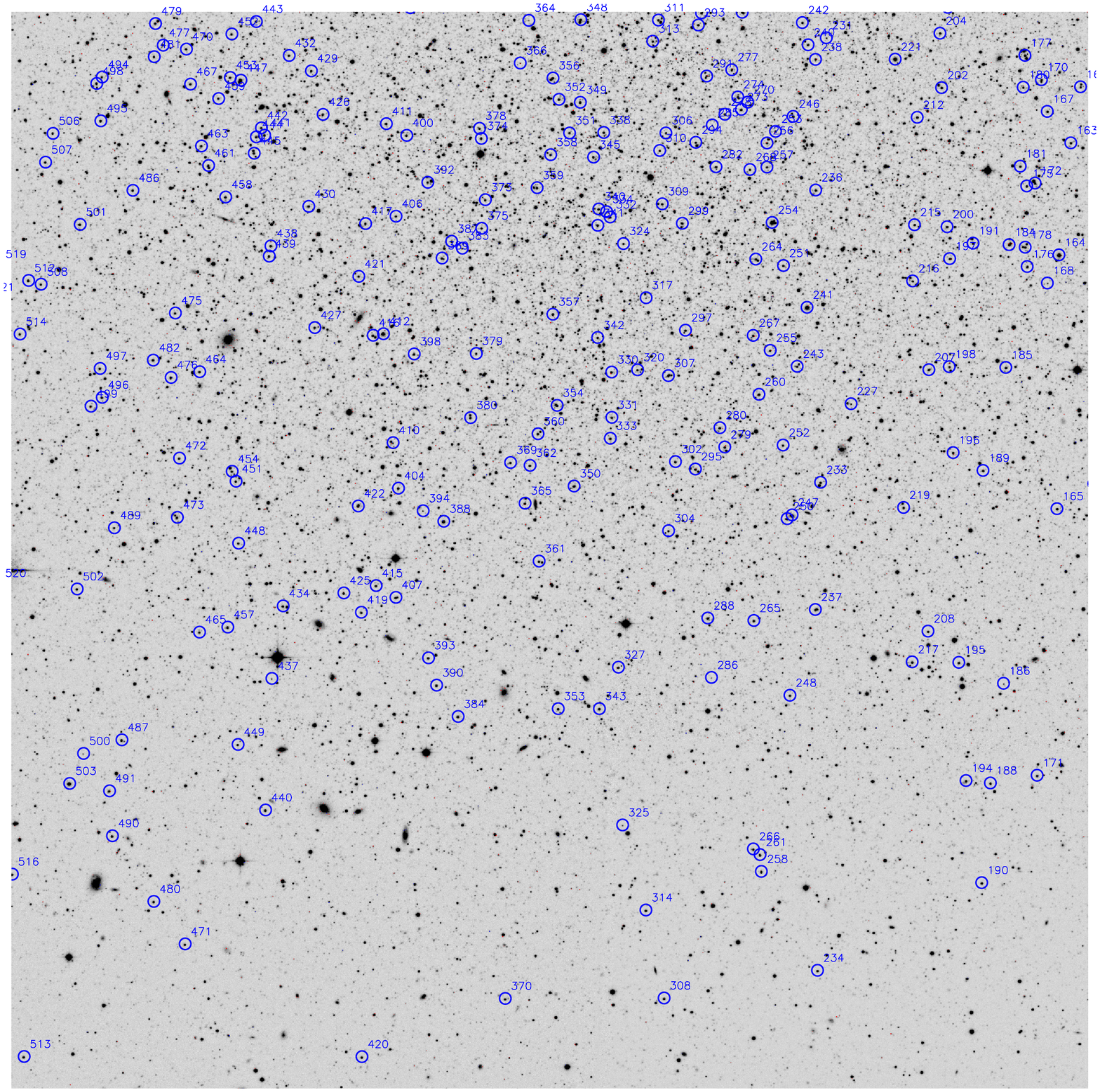}
	\caption{Finding chart for the quadrant 28 of the Mosaic of Sculptor (\ref{fig:mosaic}). 
		North is up and east to the left. The labelled numbers correspond to the numerical 
		suffix from our assigned names (scl-CEMV+suffix).The other quadrants are in the electronic edition.}
\label{fig:finding_chart}
\end{figure*}

Fig.~\ref{fig:mosaic} displays a mosaic centred on Sculptor dSph, divided in 56 quadrants.
In the electronic version we show finding charts for each quadrant in which we found 
variable stars. In this way, we make available the finding charts for the whole sample 
of variable stars detected. Fig.~\ref{fig:finding_chart} shows the finding chart for the
28 quadrant of the mosaic of Sculptor (Fig.~\ref{fig:mosaic}). The labelled numbers are 
those belonging to the numerical suffix from our assigned names (scl-CEMV+suffix).

%



\bsp	
\label{lastpage}
\end{document}